\documentclass[12pt]{iopart}
\usepackage{epsfig}
\usepackage{graphicx}

\newcommand{\here}{\makebox(0,0)}

\newcommand{\bsigma}{\mbox{\boldmath $\sigma$}}

\newcommand{\bss}{\mbox{\boldmath $s$}}

\newcommand{\bx}{\mbox{\boldmath $x$}}
\newcommand{\by}{\mbox{\boldmath $y$}}

\newcommand{\bu}{\mbox{\boldmath $u$}}
\newcommand{\bv}{\mbox{\boldmath $v$}}

\newcommand{\bQ}{\mbox{\boldmath $Q$}}
\newcommand{\bT}{\mbox{\boldmath $T$}}

\newcommand{\R}{{\rm I\!R}}
\newcommand{\room}{\rule[-0.3cm]{0cm}{0.8cm}}
\newcommand{\order}{ {\cal O}}

\newcommand{\arctanh}{ ~{\rm arctanh} }
\newcommand{\be}{\begin{equation}}
\newcommand{\ee}{\end{equation}}
\newcommand{\bd}{\begin{displaymath}}
\newcommand{\ed}{\end{displaymath}}
\newcommand{\bra}{\langle}
\newcommand{\ket}{\rangle}
\newcommand{\bigbra}{\left\langle}
\newcommand{\bigket}{\right\rangle}
\newcommand{\vsp}{\vspace*{3mm}}

\begin{document}

\title[Ising spin models on `small world' lattices]{Replicated transfer matrix analysis of Ising spin models on `small world' lattices}
\author{T Nikoletopoulos$^\dag$, A C C Coolen$^\dag$,  I P\'{e}rez Castillo$^\ddag$, N S Skantzos$^\S$,
J P L Hatchett$^\dag$ and B Wemmenhove$^\P$}

\address{\dag ~ Department of Mathematics, King's College London, The Strand,
London WC2R 2LS, United Kingdom}

\address{\ddag ~ Institute for Theoretical Physics, Celestijnenlaan 200D,
Katholieke Universiteit Leuven, B-3001 Belgium}

\address{\S ~ Departament de Fisica Fonamental, Facultat de Fisica, Universitat
de Barcelona, 08028 Barcelona, Spain}

\address{\P ~Institute for
Theoretical Physics, University of Amsterdam, Valckenierstraat 65,
1018 XE Amsterdam, The Netherlands}

\begin{abstract}
We calculate equilibrium solutions for Ising spin models on `small
world' lattices, which are constructed by super-imposing random
and sparse Poissonian graphs with finite average connectivity $c$
onto a one-dimensional ring. The nearest neighbour bonds along the
ring are ferromagnetic, whereas those corresponding to the
Poisonnian graph are allowed to be random. Our models thus
generally contain quenched connectivity and bond disorder. Within
the replica formalism, calculating the disorder-averaged free
energy requires the diagonalization of replicated transfer
matrices. In addition to developing the general replica symmetric
theory,  we derive phase diagrams and calculate effective field
distributions for two specific cases: that of uniform sparse
long-range bonds (i.e. `small world' magnets), and that of $\pm J$
random sparse long-range bonds (i.e. `small world' spin-glasses).
\end{abstract}

\pacs{75.10.Nr, 05.20.-y, 64.60.Cn} \ead{\tt
theodore@mth.kcl.ac.uk, tcoolen@mth.kcl.ac.uk,
isaac.perez@fys.kuleuven.ac.be,
 nikos@ffn.ub.es, hatchett@mth.kcl.ac.uk, wemmenho@science.uva.nl}


\section{Introduction}

The concept of `small world' networks \cite{watts-strogarz98} was
introduced as an attempt to capture and study nontrivial features
observed in realistic biological, technological and social
networks. The key idea is to generate a structure  which
interpolates between a regular finite dimensional one and a sparse
random long-range one. In a typical construction one starts with a
locally regular network, e.g. a ring, where each site is connected
to its $2k$ nearest neighbours, and subsequently `re-wires'
randomly with a probability $p$  those local connections, creating
long range shortcuts. One characteristic of networks of this type,
the so called `small world effect', is that even for small $p$
the average minimal path length between two sites can still be
very small. Small world networks have also attracted a lot of
attention in physics
\cite{monasson99,barrat-weigt00,gitterman00,kim01,herrero02}.  A
study of an Ising model \cite{barrat-weigt00} on a small world
network  showed that  even for very small re-wiring probability
$p$ there exists a ferromagnetic phase transition at finite
temperature, which is absent in the purely one-dimensional model.
From a statistical mechanics point of view, the `small world
effect' can thus be thought of as inducing global order in a
low-dimensional  system by adding a small number of long range
connections.

\begin{figure}[t]
\vspace*{-9mm}
\begin{center}{\hspace*{27mm}
\epsfxsize=105mm\epsfbox{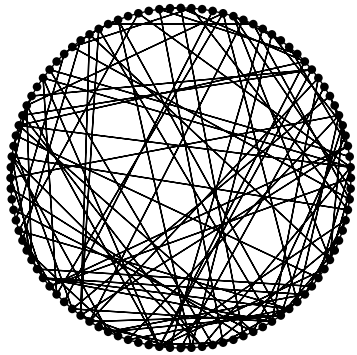} }\end{center}
\vspace*{-12mm} \caption{Example of a `small world' lattice
connecting $N$ spins (indicated by $\bullet$) in the sense of our
present models. Nearest neighbour interactions along a
one-dimensional ring are combined with sparse long-range
interactions in the form of a random Poissonian graph, with small
average connectivity $c$. Here $N=100$ and $c=2$.}
\label{fig:smallworld}
\end{figure}

In this paper we study a family of Ising spin models defined on a
lattice which, although not identical to the construction of
\cite{watts-strogarz98}, shares with the latter  the `small world'
characteristics of a regular short-range structure in coexistence
with a sparse and random range-free one.
 Our lattice consists
of a one-dimensional ring of $N$ sites occupied by Ising spins,
where each site is firstly connected to its nearest neighbours,
and secondly via a random Poissonian graph in which the average
number of connections per site  $c$ remains finite in the
thermodynamic limit $N\to\infty$. See e.g. figure
\ref{fig:smallworld}. This architecture was also studied in
\cite{gitterman00}; however, the author of \cite{gitterman00} did
not attempt to calculate transition lines or phase diagrams
analytically, but resorted to an annealed approximation. We take
the bonds between nearest neighbours on the ring to be
ferromagnetic, whereas the long-range bonds can
 be random. Our motivation for studying this structure is twofold.
Firstly, our model is a combination of a one-dimensional system
and a random finitely connected graph so its mathematical analysis
is nontrivial yet feasible, requiring an interesting mixture of
mathematical tools developed for one-dimensional models (the
transfer matrix technique) and those developed for finite
connectivity models. Finite connectivity techniques have been
applied to many areas, such as error correcting codes
\cite{murayama00,nakamura,nishimori}, theoretical computer science
\cite{kirkpatrick-selman94,monasson-zecchina98,monasson-zecchina982,monasson-zecchina99},
neural networks \cite{wemmenhove-coolen03,perez-skantzos03}, and
spin glasses
\cite{viana-bray85,kanter-sompo87,mezard-parisi87,wong-sherrington88,monasson98,mezard-parisi01,parisi-tria02}.
In the later field, research has been triggered by the desire to
develop solvable spin-glass models which are  closer to real
finite-dimensional systems than the fully connected spin-glass
model of \cite{SK}. The RKKY interactions of metallic spin-glasses
can be regarded as a combination of short-range ferromagnetic
bonds and long-range random ones. Thus there is a case for
regarding our `small world' networks, which are defined with a
similar structure of spin-interactions, as perhaps more realistic
than finite connectivity spin-glass models. This is our second
motivation.

At a technical level,  upon calculating for the present class of
models an expression for the disorder-averaged free energy per
spin using the replica formalism, one is led to a replicated
transfer matrix structure embedded within a mean-field
finite-connectivity type calculation. Solving our models thereby
boils down to the diagonalization of specific $2^n\times 2^n$
replicated transfer matrices, in the limit $n\to 0$. We show how
this can be done within the replica symmetric ansatz, and we use
the result to calculate fully explicit expressions for transition
lines, phase diagrams, and effective field distributions.
Numerical simulations show perfect agreement with our theory.

Upon completion of this study,  a preprint was communicated
\cite{prep} in which the authors also aim to solve an Ising model
on a small world lattice explicitly. However, both their methods
(combinatorics rather than replicated transfer matrices) and model
definitions (regular rather than random long-range sparse
connectivity) differ from those in the present paper.

\section{Definitions and replica analysis}

Our model is a system of $N$ interacting Ising spins
$\sigma_{i}\in\{-1,1\}$, in thermal equilibrium at inverse
temperature $\beta=1/T$, described by the following Hamiltonian
(defined with $\sigma_{N+1}\equiv \sigma_1$):
\be
  H=-J_{0}\sum_{i}\sigma_{i}\sigma_{i+1}
             -\frac{1}{c}\sum_{i<j}J_{ij}c_{ij}\sigma_{i}\sigma_{j}
             \label{eq:Hamiltonian}
\ee
 The (long-range) couplings $J_{ij}\in\R$ are independent
 identically distributed
random variables, drawn from some distribution $p(J_{ij})$. We
will abbreviate $\int\!dJ~p(J)g(J)=\bra g(J)\ket_J$. The variables
$c_{ij}\in\{0,1\}$ specify whether a long-range bond between sites
$i$ and $j$ is present ($c_{ij}=1$) or absent ($c_{ij}=0$); they
are also independent random variables, drawn according to
\be
  p(c_{ij})=\frac{c}{N}\delta_{c_{ij},1}+\left(1-\frac{c}{N}\right)\delta_{c_{ij},0}
  \label{eq:connectivity}
\ee This choice leads in the thermodynamic limit $N\to\infty$ to a
number of long range connections  per site distributed according
to a Poisson law with mean $c$. The number $c$ will remain finite
in the thermodynamic limit  $N\to\infty$. Equilibrium ensemble
averages, with the conventional Boltzmann measure for spin
configurations, will simply be denoted by $\bra \ldots \ket$. The
dilution variables and long-range bonds $\{c_{ij},J_{ij}\}$
represent quenched disorder, with respect to which all macroscopic
thermodynamic quantities are assumed to be self averaging in the
limit $N\to\infty$. We will write disorder averages as
$\overline{[\ldots]}$. \vsp

 We calculate the
asymptotic disorder-averaged free energy per spin in the
conventional manner using the replica formalism, which is based on
the identity
\be
  \overline{f}=-\lim_{N\to\infty}\frac{1}{\beta N}\lim_{n\to
  0}\frac{1}{n}\log\overline{Z^n}
  ~~~~~~~~Z=\sum_{\sigma_1,\ldots,\sigma_N} e^{-\beta H}
\ee As usual, the disorder average is performed for integer $n$,
the order of the  limits $N\to\infty$ and $n\to 0$ is reversed,
and the final result is extended to non-integer $n$ to allow the
limit $n\to 0$ to be taken. We will denote $n$-replicated spins by
$\bsigma_{i}=(\sigma_{1}^{i},\ldots,\sigma_{n}^{i})$, and
abbreviate $\{\bsigma\}=\{\bsigma_{1},\ldots,\bsigma_{N}\}$.
Averaging $Z^n$ over the disorder now gives
\begin{eqnarray}
\overline{ Z^{n}}&=&
     \sum_{\{\bsigma\}}e^{\beta J_{0}\sum_{i}\bsigma_{i}\cdot\bsigma_{i+1}}
       \overline{\left[ e^{\frac{\beta}{c}\sum_{i<j}J_{ij}c_{ij}\bsigma_{i}\cdot\bsigma_{j}}\right]} \nonumber \\
   &=&
     \sum_{\{\bsigma\}}e^{\beta J_{0}\sum_{i}\bsigma_{i}\cdot\bsigma_{i+1}+\frac{c}{N}
     \sum_{i<j}\bra e^{\frac{\beta
J}{c}\bsigma_{i}\cdot\bsigma_{j}}-1\ket_{J}+\order(N^0)}
\end{eqnarray}
This expression can be transformed into an integral to be
calculated by steepest descent as $N\to\infty$, via the
introduction of the order parameter distribution
$P(\bss;\{\bsigma\})=N^{-1}\sum_{i}\delta_{\bss,\bsigma_{i}}$.
This latter observable gives the fraction of sites with a given
configuration $\bss\in\{-1,1\}^{n}$ of replicated spin variables
\cite{monasson98}.
 In particular we find
\be
   \overline{Z^{n}}=
   \int\!\prod_{\bss}\left[\rmd P(\bss)\rmd\hat{P}(\bss)\right]
     e^{N\Psi[\{P,\hat{P}\}]}
\label{eq:saddle_point_form}
 \ee
 $~$\vspace*{-3mm}
\begin{eqnarray}
  \Psi[\{P,\hat{P}\}]&=&
  \rmi\sum_{\bss}P(\bss)\hat{P}(\bss)+\frac{1}{2}c\sum_{\bss,\bss^\prime}P(\bss)P(\bss^\prime)
            \bra e^{\frac{\beta J}{c}\bss\cdot\bss^\prime}-1\ket_{J}
   \nonumber  \\
  &&+\frac{1}{N}\log\sum_{\{\bsigma\}}e^{\beta J_{0}\sum_{i}\bsigma_{i}\cdot\bsigma_{i+1}-
        \rmi\sum_i\hat{P}(\bsigma_{i})}
        +\order(\frac{\log
        N}{N})
\end{eqnarray}
In the limit $N\to\infty$ the integral
(\ref{eq:saddle_point_form}) is dominated by the stationary point
of $\Psi[\{P,\hat{P}\}]$. Working out the equations $\partial
\Psi/\partial P(\bss)=\partial \Psi/\partial \hat{P}(\bss)=0$, for
all $\bss$, gives respectively
\begin{eqnarray}
  \hat{P}(\bss)&=&\rmi c\sum_{\bss^\prime}P(\bss^\prime)\bra e^{\frac{\beta J}{c}\bss\cdot\bss^\prime}-1\ket_{J}
  \label{eq:SPE1}
  \\
  P(\bss)&=&\lim_{N\to\infty}
  \frac{\sum_{\{\bsigma\}}e^{\beta J_{0}\sum_{i}\bsigma_{i}\cdot\bsigma_{i+1}-
        \rmi\sum_i\hat{P}(\bsigma_{i})}\left(\frac{1}{N}\sum_{j}\delta_{\bss,\bsigma_{j}}\right)}
                {\sum_{\{\bsigma\}}e^{\beta J_{0}\sum_{i}\bsigma_{i}\cdot\bsigma_{i+1}-
        \rmi\sum_i\hat{P}(\bsigma_{i})}}
        \label{eq:SPE2}
\end{eqnarray}
We use (\ref{eq:SPE1}) to eliminate the conjugate order
parameters.  To proceed further we define a $2^{n}\times 2^{n}$
transfer matrix $\bT[P]$ and a diagonal $2^{n}\times 2^{n}$ matrix
$\bQ[\bss]$, with entries
\begin{eqnarray}
  T_{\bsigma,\bsigma^\prime}[P]&=& e^{\beta J_{0}\bsigma\cdot \bsigma^\prime +
       c\sum_{\bss}P(\bss)\bra e^{\frac{\beta J}{c}\bsigma\cdot \bss}-1\ket_{J} }
       \label{eq:RTM}
\\
Q_{\bsigma,\bsigma^\prime}[\bss]&=&\delta_{\bss,\bsigma}\delta_{\bsigma,\bsigma^\prime}
\label{eq:defineQ}
\end{eqnarray}
These definitions allow us to write  the replicated spin
summations as traces of matrix products, so that, upon inserting
(\ref{eq:SPE1}) into (\ref{eq:SPE2}), the latter becomes
 \begin{eqnarray}
 \hspace*{-10mm}
&&
P(\bss)=\lim_{N\to\infty}\frac{1}{N}\sum_{j}\frac{\tr(\bT^{j-1}[P]~\bQ[\bss]~\bT^{N-j+1}[P])}{\tr(\bT^{N}[P])}
 =\lim_{N\to\infty}\frac{\tr(\bQ[\bss]\bT^{N}[P])}{\tr(\bT^{N}[P])}
 \label{eq:SPEcombined}
\end{eqnarray}
Only contributions from the largest eigenvalue $\lambda(n;P)$ of
(\ref{eq:RTM}) survive the  limit $N\to\infty$. If we denote by
$\bv[P]$ and $\bu[P]$ the associated left and right eigenvectors,
with components $v_{\bsigma}[P]$ and $u_{\bsigma}[P]$, we arrive
at the following expressions for (\ref{eq:SPEcombined}) and the
free energy:
\begin{eqnarray}
  P(\bss)&=&\frac{v_{\bss}[P]~
  u_{\bss}[P]}{\sum_{\bss^\prime}v_{\bss^\prime}[P]~
  u_{\bss^\prime}[P]}
\label{eq:SPE_RSB}
\\
  \overline{f}&=&\lim_{n\to 0}\frac{1}{n\beta}\bigg\{
    \frac{1}{2}c\sum_{\bss,\bss'}P(\bss)P(\bss')\bra e^{\frac{\beta J}{c}\bss\cdot\bss'}-1\ket_{J}
      -\log\lambda(n;P)\bigg\}
\label{eq:f_RSB}
\end{eqnarray}
In order to proceed we next need to solve the eigenvalue problem
for the replicated transfer matrix $\bT[P]$ as defined in
(\ref{eq:RTM}). This is generally a nontrivial task, therefore  in
the remainder of our paper  we will focus on the replica symmetric
solution.

\section{Solution by diagonalization of RS replicated transfer matrix}

\subsection{Conversion to a functional eigenvalue problem}

The ergodic, or replica symmetric (RS), ansatz corresponds to the
distribution $P(\bss)$ being invariant under all permutations of
the replica labels $\{1,\ldots,n\}$. Now $P(\bss)$ can depend on
$\bss$ only via the sum $\sum_{\alpha=1}^n s_{\alpha}$, and may
thus always be written in the form
\be
  P(\bss)=\int\!\rmd h~W(h)\prod_{\alpha=1}^n \frac{e^{\beta h s_{\alpha}}}{2\cosh(\beta h)}
  \label{eq:RSansatz}
\ee where the new RS order parameter $W(h)$ is a (normalized)
distribution of effective fields $h_i$, which are defined via
$\bra\sigma_i\ket=\tanh(\beta h_i)$.
 Let us next define the following short-hand:
\be
\hspace*{-10mm}
  w(\sum_{\alpha}\sigma_{\alpha},n)=
    \sum_{\bss^\prime}P(\bss^\prime )\bra e^{\frac{\beta J}{c}\bsigma\cdot\bss^\prime}\ket_{J}=
    \int\!\rmd h~W(h)\frac{\bra e^{\beta [A(\frac{J}{c},h)\sum_{\alpha}\sigma_{\alpha}+  n B(\frac{J}{c},h)] }\ket_{J}}
                              {[2\cosh(\beta h)]^{n}}
\ee
where
\begin{eqnarray*}
A(J,z)&=&\frac{1}{2\beta}\log\left[\frac{\cosh[\beta(J+z)]}{\cosh[\beta(J-z)]}\right]=
           \frac{1}{\beta}\arctanh[\tanh(\beta J)\tanh[\beta z)]  \\
           B(J,z)&=&\frac{1}{2\beta}\log\Big[
4\cosh[\beta(J+z)]\cosh[\beta(J-z)] \Big]
\end{eqnarray*}
This allows us to write the entries of
 our transfer matrix (\ref{eq:RTM}) within the RS ansatz as
\begin{eqnarray}
  T^{\rm RS}_{\bsigma,\bsigma^\prime}&=& e^{\beta J_{0}\bsigma\cdot \bsigma^\prime +
       c w(\sum_{\alpha}\sigma_{\alpha},n)-c}
\nonumber
\\
&=& \int\!\rmd \hat{m}~R(\hat{m},n) \prod_\alpha e^{\beta
J_{0}\sigma_\alpha\sigma_\alpha^\prime
-\rmi\beta\hat{m}\sigma_{\alpha}} \label{eq:RTM2}
\end{eqnarray}
with
 \be R(\hat{m},n)=
\beta \int\!\frac{\rmd m}{2\pi}~e^{\rmi\beta\hat{m}m+ c w(m,n)-c}
\label{eq:defineR}
 \ee
 The matrix (\ref{eq:RTM2}) is formally
 identical to the replicated transfer matrix which would have been found for Ising chains with (complex) random fields $-\rmi
 \hat{m}$, distributed according to $R(\hat{m},n)$.
We exploit this equivalence, and postulate eigenvectors with the
structure as found in one-dimensional random-field models. In
particular, for the right and left eigenvectors of the largest
eigenvalue of (\ref{eq:RTM2}) we substitute a form which is a
complex extension of that  introduced in \cite{nik-coolen04}:
\begin{eqnarray}
  u_{\bsigma}^{\rm RS}&=&\int\!\rmd x_1 \rmd x_2~\Phi(x_1,x_2|n)~e^{\beta (x_1+\rmi x_2)\sum_{\alpha}\sigma_{\alpha}}
  \label{eq:right_ev}\\
  v_{\bsigma}^{\rm RS}&=&\int\!\rmd y_1 \rmd y_2~\Psi(y_1,y_2|n)~e^{\beta (y_1+\rmi y_2)\sum_{\alpha}\sigma_{\alpha}}
  \label{eq:left_ev}
\end{eqnarray}
 In one-dimensional models $x_1+\rmi x_2$ and $y_1+\rmi y_2$ would represent fields which are
propagated along the chain, according to a random Markovian map
which reflects the statistical properties of the random fields and
 nearest neighbour interactions. In the limit $N\to\infty$,
$\Phi(\ldots|n)$ and $\Psi(\ldots|n)$ give the  distributions of
these fields, invariant  under the map.  Insertion of
(\ref{eq:right_ev},\ref{eq:left_ev}) into the eigenvalue equations
(to be satisfied for every $\bsigma$), i.e. \bd
\sum_{\bsigma^\prime}T^{\rm RS}
_{\bsigma,\bsigma^\prime}u_{\bsigma^\prime}^{\rm RS} =\lambda_{\rm
RS}(n)u_{\bsigma}^{\rm RS}, ~~~~~~~ \sum_{\bsigma^\prime}v^{\rm
RS}_{\bsigma^\prime}T^{\rm
RS}_{\bsigma^\prime\!,\bsigma}=\lambda_{\rm RS}(n)v_{\bsigma}^{\rm
RS}
\ed leads to new eigenvalue problems for $\Phi$ and $\Psi$,
with $\bx=(x_1,x_2)$ and  $\by=(y_1,y_2)$:
\begin{eqnarray}
  \lambda_{\rm RS}(n)\Phi(\bx|n)&=&\int\!\rmd \bx^\prime~\Lambda_{\Phi}(\bx,\bx^\prime|n)~\Phi(\bx^\prime|n)
  \label{eq:new_ev_problem1}
\\
  \lambda_{\rm RS}(n)\Psi(\by|n)&=&\int\!\rmd \by^\prime~\Lambda_{\Psi}(\by,\by^\prime|n)~\Psi(\by^\prime|n)
  \label{eq:new_ev_problem2}
\end{eqnarray}
with the complex kernels
\begin{eqnarray}
\hspace*{-23mm}
 \Lambda_{\Phi}(\bx,\bx^\prime|n) &=&
\int\!\rmd\hat{m}~R(\hat{m},n)~e^{n\beta B(J_{0},x_1^\prime+\rmi
x_2^\prime)}~
     \delta\left(\!\begin{array}{l}
     x_1-{\rm Re}\{ A(J_0,x_1^\prime\!+\rmi x_2^\prime)-\rmi \hat{m}\}\\
     x_2-{\rm Im}\{ A(J_0,x_1^\prime\!+\rmi x_2^\prime)-\rmi
     \hat{m}\}
     \end{array}\!\right)
     \label{eq:kernel1}
\\
\hspace*{-23mm} \Lambda_{\Psi}(\by,\by^\prime|n)
 &=&
    \int\!\rmd\hat{m}~R(\hat{m},n)~e^{n\beta B(J_{0},y_1^\prime+\rmi y_2^\prime-\rmi\hat{m})}
~   \delta\left(\!\begin{array}{l}
     y_1-{\rm Re}~A(J_0,y_1^\prime\!+\rmi y_2^\prime\!-\rmi\hat{m})\\
     y_2-{\rm Im}~A(J_0,y_1^\prime\!+\rmi
     y_2^\prime\!-\rmi\hat{m})
     \end{array}\!\right)
     \label{eq:kernel2}
\end{eqnarray}
In order to find a self-consistent equation for the RS order
parameter $W(h)$ in (\ref{eq:RSansatz}) we need to inspect only
the $n\to 0$ limit of the eigenvalue problems
(\ref{eq:new_ev_problem1},\ref{eq:new_ev_problem2}). For the free
energy, however, we need to know the first two orders in $n$ of
the eigenvalue $\lambda_{\rm RS}(n)$.

\subsection{Derivation of the  RS order parameter equation}

Upon taking the limit $n\to 0$ we find the kernels
(\ref{eq:kernel1},\ref{eq:kernel2}) reducing to
\begin{eqnarray}
 \Lambda_{\Phi}(\bx,\bx^\prime|0) &=&
\int\!\rmd\hat{m}~R(\hat{m},0)~
     \delta\left(\!\begin{array}{l}
     x_1-{\rm Re}\{A(J_0,x_1^\prime\!+\rmi x_2^\prime)-\rmi\hat{m}\}\\
     x_2-{\rm Im}\{A(J_0,x_1^\prime\!+\rmi
     x_2^\prime)-\rmi\hat{m}\}
     \end{array}\!\right)
\\
\Lambda_{\Psi}(\by,\by^\prime|0)
 &=&
    \int\!\rmd\hat{m}~R(\hat{m},0)
~   \delta\left(\!\begin{array}{l}
     y_1-{\rm Re}~A(J_0,y_1^\prime\!+\rmi y_2^\prime\!-\rmi\hat{m})\\
     y_2-{\rm Im}~A(J_0,y_1^\prime\!+\rmi
     y_2^\prime\!-\rmi\hat{m})
     \end{array}\!\right)
\end{eqnarray}
Subsequent integration of the eigenvalue equations
(\ref{eq:new_ev_problem1},\ref{eq:new_ev_problem2}) over $\bx$ and
$\by$, for $n\to 0$ and using $\int\!\rmd\bx~\Phi(\bx|0)>0$ and
$\int\!\rmd\by~\Psi(\by|0)>0$ (since $\Phi$ and $\Psi$ represent
field distributions), leads in both cases to the following
identity
\be
  \lambda_{\rm RS}(0) =
\int\!\rmd\hat{m}~R(\hat{m},0)= e^{ c w(0,0)-c}=1
 \ee
which is indeed necessary for the $n\to 0$ limit in
(\ref{eq:f_RSB}) to exist. Knowing $\lambda_{\rm RS}(0)$, we may
in turn write the two $n\to 0$ eigenvalue problems as
\begin{eqnarray*}
\hspace*{-10mm}
 \Phi(\bx|0)
  &=& \!\int\!\!\rmd\bx^\prime~ \Phi(\bx^\prime|0)
  \int\!\frac{\rmd z\rmd m}{2\pi} e^{\rmi zm+ c w(m,0)-c}~
     \delta\!\left(\!\begin{array}{l}
     x_1\!-{\rm Re}\{ A(J_0,x_1^\prime\!+\rmi x_2^\prime)-\frac{\rmi z}{\beta}\}\\
     x_2\!-{\rm Im}\{ A(J_0,x_1^\prime\!+\rmi x_2^\prime)-\frac{\rmi
     z}{\beta}\}
     \end{array}\!\right)
\\
\hspace*{-10mm}
 \Psi(\by|0)
 &=&
 \!\int\!\!\rmd \by^\prime ~\Psi(\by^\prime|0)
    \int\!\frac{\rmd z\rmd m}{2\pi} e^{\rmi zm+ c w(m,0)-c}
~   \delta\!\left(\!\begin{array}{l}
     y_1\!-{\rm Re}~A(J_0,y_1^\prime\!+\rmi y_2^\prime\!-\frac{\rmi z}{\beta})\\
     y_2\!-{\rm Im}~A(J_0,y_1^\prime\!+\rmi
     y_2^\prime\!-\frac{\rmi z}{\beta})
     \end{array}\!\right)
\end{eqnarray*}
We now expand $e^{cw(m,0)}$ in powers of $c$, using $w(m,0)=
\int\!\rmd h~W(h)\bra e^{\beta m A(\frac{J}{c},h) }\ket_J$:
\begin{eqnarray}
&&\hspace*{-15mm}
 \int\!\frac{\rmd z\rmd m}{2\pi} e^{\rmi zm+ c w(m,0)-c}\delta\left(\room\ldots\right)=
\sum_{k\geq 0}\frac{ e^{-c} c^k}{k!}\int\!\frac{\rmd z\rmd
m}{2\pi}~ e^{\rmi
 zm} w^k(m,0)\delta\left(\room\!\ldots\right)
 \nonumber
 \\
 &=&
 \sum_{k\geq 0}\frac{ e^{-c} c^k}{k!}
 \int\!\prod_{\ell=1}^k\left[ \rmd h_\ell\rmd J_\ell
W(h_\ell)p(J_\ell) \right] \int\!\frac{\rmd z\rmd m}{2\pi}
\delta\left(\room\!\ldots\right) e^{\rmi
 zm+\beta m \sum_{\ell=1}^k
A(\frac{J_\ell}{c},h_\ell) } \nonumber
 \end{eqnarray}
We assume our functions to decay to zero sufficiently fast away
from the origin to allow us to shift the $z$-integration in the
complex plane according to $z\to z +i\beta \sum_{\ell=1}^k
A(\frac{J_\ell}{c},h_\ell)$. This enables us to perform the
integral over $m$, giving $2\pi\delta(z)$, and write
\begin{eqnarray}
 \Phi(\bx|0)
  &=& \!\int\!\!\rmd\bx^\prime~ \Phi(\bx^\prime|0)
 \sum_{k\geq 0}\frac{ e^{-c} c^k}{k!}
 \int\!\prod_{\ell=1}^k\left[ \rmd h_\ell\rmd J_\ell
W(h_\ell)p(J_\ell) \right] \nonumber
\\
&&\times
     \delta\!\left(\!\begin{array}{l}
     x_1\!-{\rm Re}\{ A(J_0,x_1^\prime\!+\rmi x_2^\prime)+ \sum_{\ell=1}^k
A(\frac{J_\ell}{c},h_\ell)\}\\
     x_2\!-{\rm Im}\{ A(J_0,x_1^\prime\!+\rmi x_2^\prime)+ \sum_{\ell=1}^k
A(\frac{J_\ell}{c},h_\ell)\}
     \end{array}\!\right)
\label{eq:Phi_eqn_aftershift}
\\
 \Psi(\by|0)
 &=&
 \!\int\!\!\rmd \by^\prime ~\Psi(\by^\prime|0)
\sum_{k\geq 0}\frac{ e^{-c} c^k}{k!}
 \int\!\prod_{\ell=1}^k\left[ \rmd h_\ell\rmd J_\ell
W(h_\ell)p(J_\ell) \right] \nonumber
\\
&&\times    \delta\!\left(\!\begin{array}{l}
     y_1\!-{\rm Re}~A(J_0,y_1^\prime\!+\rmi y_2^\prime\!+\sum_{\ell=1}^k
A(\frac{J_\ell}{c},h_\ell))\\
     y_2\!-{\rm Im}~A(J_0,y_1^\prime\!+\rmi
     y_2^\prime\!+ \sum_{\ell=1}^k
A(\frac{J_\ell}{c},h_\ell))
     \end{array}\!\right)
\label{eq:Psi_eqn_aftershift}
\end{eqnarray}
These eigenvalue equations allow for solutions describing the
expected manifestly real-valued fields, i.e.
$\Phi(x_1,x_2|0)=\delta(x_2)\Phi(x_1)$ and
$\Psi(y_1,y_2|0)=\delta(y_2)\Psi(y_1)$, where
\begin{eqnarray}
 \Phi(x)
  &=& \int\!\rmd x^\prime~ \Phi(x^\prime)
 \sum_{k\geq 0}\frac{ e^{-c} c^k}{k!}
 \int\!\prod_{\ell=1}^k\left[ \rmd h_\ell\rmd J_\ell
W(h_\ell)p(J_\ell) \right] \nonumber
\\
&&\hspace*{20mm}
\times
     \delta[
     x- A(J_0,x^\prime)- \sum_{\ell=1}^k
A(\frac{J_\ell}{c},h_\ell) ] \label{eq:PHI}
\\
\Psi(y)
 &=&
 \int\!\rmd y^\prime ~\Psi(y^\prime)
\sum_{k\geq 0}\frac{ e^{-c} c^k}{k!}
 \int\!\prod_{\ell=1}^k\left[ \rmd h_\ell\rmd J_\ell
W(h_\ell)p(J_\ell) \right] \nonumber
\\
&&\hspace*{20mm} \times \delta[y-A(J_0,y^\prime\!+\sum_{\ell=1}^k
A(\frac{J_\ell}{c},h_\ell))] \label{eq:PSI}
\end{eqnarray}
We may always normalize our solutions such that $\int\!\rmd
x~\Phi(x)=\int\!\rmd y~\Psi(y)=1$, which will be assumed from now
on. Finally, for RS states
(\ref{eq:RSansatz},\ref{eq:right_ev},\ref{eq:left_ev}) with
$\Phi(x_1,x_2|0)=\delta(x_2)\Phi(x_1)$ and
$\Psi(y_1,y_2|0)=\delta(y_2)\Psi(y_1)$ we find the
self-consistency equation (\ref{eq:SPE_RSB}) reducing to
\be
\int\!\rmd h~W(h) \frac{e^{\beta h \sum_\alpha
s_{\alpha}}}{[2\cosh(\beta h)]^n}
  =\frac{\int\!\rmd x \rmd y~\Phi(x)\Psi(y) e^{\beta (x+y)\sum_{\alpha}s_{\alpha}}}
 {\int\!\rmd x \rmd y~\Phi(x)\Psi(y)[2\cosh[\beta(x+y)]]^n}
\ee From this we obtain for $n=0$ the simple expression
\be
  W(h)=\int\!\mathrm{d}x\mathrm{d}y~\Phi(x)\Psi(y)\delta[h-x-y]
  \label{eq:W}
\ee The final trio of coupled equations
(\ref{eq:PHI},\ref{eq:PSI},\ref{eq:W}), to be solved
simultaneously, determine the effective field distributions
$W(h),\Phi(x),\Psi(y)$. The latter functions play the role of RS
order parameters in our theory.

\subsection{Derivation of the  RS free energy}

We now turn to the evaluation of the free energy per spin
 in RS ansatz, by inserting (\ref{eq:RSansatz}) into (\ref{eq:f_RSB}):
\begin{eqnarray}
  \overline{f}_{\rm RS}&=&
 \lim_{n\to 0}\frac{c}{2n\beta} \bigbra
 \int\!\rmd h \rmd h^\prime~W(h)W(h^\prime)
\left[\frac{\sum_{\sigma,\sigma^\prime} e^{\beta(h \sigma+
h^\prime \sigma^\prime+\frac{J}{c}\sigma
\sigma^\prime)}}{4\cosh(\beta h) \cosh(\beta
h^\prime)}\right]^n-1\bigket_{\!J} \nonumber
\\
&&\hspace*{40mm}
      -\lim_{n\to 0}\frac{1}{n\beta}\log\lambda_{\rm RS}(n)
      \nonumber
    \\
&=&
 \frac{c}{2\beta}
 \int\!\rmd h \rmd h^\prime~W(h)W(h^\prime)\bra
 \log
[1+\tanh(\frac{\beta J}{c})\tanh(\beta h)\tanh(\beta h^\prime)]
 \ket_{J} \nonumber
\\
&&\hspace*{20mm} + \frac{c}{2\beta}
 \bra
 \log
\cosh(\frac{\beta J}{c}) \ket_{\!J}
      -\lim_{n\to 0}\frac{1}{n\beta}\log\lambda_{\rm RS}(n)
      \label{eq:f_RS1}
\end{eqnarray}
In evaluating $\lambda_{\rm RS}(n)$ we will need the following two
identities:
\begin{eqnarray}
\hspace*{-20mm} w(0,n)&=&1+n\int\! \rmd h~W(h)\left\{ \beta \bra
B(\frac{J}{c},h)\ket_J-\log[2\cosh(\beta h)]\right\}+\order(n^2)
\\
\hspace*{-20mm}
 \int\!\rmd \hat{m}~R(\hat{m},n)&=&
 e^{c w(0,n)-c}\nonumber\\
\hspace*{-20mm}
 &=&1+ nc\int\! \rmd h~W(h)\left\{ \beta \bra
B(\frac{J}{c},h)\ket_J-\log[2\cosh(\beta h)]\right\}+\order(n^2)
\end{eqnarray}
We can now obtain an expression for $\lambda_{\rm RS}(n)$ by
integrating both sides of equation (\ref{eq:new_ev_problem1}):
\begin{eqnarray}
\lim_{n\to 0} \frac{1}{n}\log \lambda_{\rm RS}(n)&=&\lim_{n\to 0}
\frac{1}{n}\log\left\{\frac{\int\!\rmd\bx \rmd
  \bx^\prime~\Lambda_{\Phi}(\bx,\bx^\prime|n)~\Phi(\bx^\prime|n)}{\int\!\rmd\bx~\Phi(\bx|n)}
  \right\}\nonumber
  \\
&=& \lim_{n\to 0}\frac{1}{n}\log\left\{\frac{\int\! \rmd
  \bx~\int\!\rmd \hat{m}~R(\hat{m},n)
e^{n\beta B(J_{0},x_1+\rmi
x_2)}~\Phi(\bx|n)}{\int\!\rmd\bx~\Phi(\bx|n)} \right\} \nonumber
  \\
&=&
  c\int\! \rmd h~W(h)\left\{ \beta \bra
B(\frac{J}{c},h)\ket_J-\log[2\cosh(\beta h)]\right\}\nonumber \\
&& + \beta\int\! \rmd
  x~\Phi(x)
B(J_{0},x)
  \end{eqnarray}
Insertion into (\ref{eq:f_RS1}) subsequently gives, after some
simple manipulations:
\begin{eqnarray}
\overline{f}_{\rm RS}&=&
 \frac{c}{2\beta}
 \int\!\rmd h \rmd h^\prime~W(h)W(h^\prime)\bra
 \log
[1+\tanh(\frac{\beta J}{c})\tanh(\beta h)\tanh(\beta h^\prime)]
 \ket_{J} \nonumber
\\
&& -  \frac{c}{2\beta}\int\! \rmd h~W(h)\bra \log[1-
\tanh^2(\frac{\beta J}{c})\tanh^2(\beta h) ] \ket_J \nonumber
\\ && - \frac{c}{2\beta}
 \bra
 \log
\cosh(\frac{\beta J}{c}) \ket_{J} \nonumber
\\ &&  - \frac{1}{2\beta}\int\! \rmd
  x~\Phi(x)
\log\Big[ 4\cosh[\beta(J_0+x)]\cosh[\beta(J_0-x)] \Big]
      \label{eq:f_RS_final}
\end{eqnarray}

\section{General mathematical properties of our RS equations}

\subsection{Special limits}

Let us first confirm that for the two extreme choices
$p(J)=\delta(J)$ (equivalently $c=0$, where we return to a simple
ferromagnetic ring) and $J_0=0$ (where we retain only the sparse
Poissonian graph) we recover from our  present full RS theory the
appropriate simpler equations known from earlier literature, to
serve as benchmark tests.

We will repeatedly benefit from the property $A(0,z)=0$.  For
$p(J)=\delta(J)$ our RS order parameter equations
(\ref{eq:PHI},\ref{eq:PSI}) simply give $\Phi(x)=\Psi(x)$ for all
$x$, with
\be
 \Phi(x)
  = \int\!\rmd x^\prime~ \Phi(x^\prime)\delta[  x- A(J_0,x^\prime)]
\label{eq:Phi_ring}
 \ee
  Any solution $\Phi$ of this equation can
have support only at
 periodic points of the non-linear map  $x\to A(J_0,x)$. Upon writing $y=\tanh(\beta J_0)\tanh(\beta x)$
  this map can be simplified to $y\to \tanh(\beta J_0)y$, so the
  only periodic point is the trivial fixed-point $x=0$.
It follows that the only normalized solution of
(\ref{eq:Phi_ring}) is $\Phi(x)=\delta(x)$, with which we obtain
the trivial effective field distribution $W(h)=\delta(h)$, as we
should, and
 the correct free energy of the one-dimensional Ising model:
\begin{eqnarray}
\overline{f}_{\rm RS}&=&
   - \frac{1}{\beta}\log[2\cosh(\beta J_0)]
\end{eqnarray}
For $J_0=0$, on the other hand, our order parameter equations tell
us that $\Psi(y)=\delta(y)$ and hence $W(h)=\Phi(h)$. Now we
recover the familiar equation for $W(h)$ corresponding to purely
Poissonian finite connectivity spin-glass models, which can be
found in e.g. \cite{mezard-parisi87}:
\begin{eqnarray}
 W(h)  &=&
 \sum_{k\geq 0}\frac{ e^{-c} c^k}{k!}
 \int\!\prod_{\ell=1}^k\left[ \rmd h_\ell\rmd J_\ell
W(h_\ell)p(J_\ell) \right] \nonumber \\ &&\times \delta\left[
     h- \frac{1}{\beta}\sum_{\ell=1}^k
\arctanh[\tanh(\frac{\beta J_\ell}{c})\tanh(\beta h_\ell)]\right]
\end{eqnarray}

\subsection{Scalar RS observables}

The conventional RS scalar observables such as the magnetization
$m$ and the spin-glass order parameter $q$ can be obtained as
always as integrals over the effective field distribution $W(h)$.
The starting point are the usual replica identities
\begin{eqnarray*}
  \overline{\bra\sigma_{i}\ket}&=&
   \lim_{n\to 0}\frac{1}{n}\sum_{\alpha}\sum_{\{\bsigma\}}\sigma_{i}^{\alpha}
       \overline{\left[\prod_{\gamma=1}^{n}e^{-\beta
       H(\bsigma^{\gamma})}\right]}
\\
  \overline{\bra\sigma_{i}\ket^{2}}
  &=&\lim_{n\to 0}\frac{1}{n(n-1)}\sum_{\alpha\neq \beta}\sum_{\{\bsigma\}} \sigma_{i}^{\alpha}\sigma_{i}^{\beta}
       \overline{\left[
       \prod_{\gamma=1}^{n}e^{-\beta H(\bsigma^{\gamma})}\right]}
\end{eqnarray*}
 Upon repeating all those manipulations which we followed
previously in calculating the RS free energy, one arrives at the
familiar expressions for
$m=\lim_{N\to\infty}N^{-1}\sum_i\overline{\bra \sigma_i\ket}$ and
$q=\lim_{N\to\infty}N^{-1}\sum_i\overline{\bra \sigma_i\ket^2}$:
\be
  m=\int\mathrm{d}h~W(h)\tanh(\beta h)
~~~~~~~~
  q=\int\mathrm{d}h~W(h)\tanh^{2}(\beta h)
  \label{eq:q_and_m}
\ee

\subsection{Continuous bifurcations away from the paramagnetic state}

Inspection of the three coupled equations
(\ref{eq:PHI},\ref{eq:PSI},\ref{eq:W}) shows that, due to
$A(J,0)=0$, the paramagnetic state
$\Phi(h)=\Psi(h)=W(h)=\delta(h)$ is always a solution. For
$\beta=0$ (high temperatures), where $A(J,z)=0$ for any $(J,z)$,
it is the only solution. As always in finite connectivity theory,
we can find continuous bifurcations away  from the paramagnetic
solution (upon lowering the temperature) by assuming that close to
the transition the effective fields are very narrowly distributed
around zero. This enables us to expand our equations in moments of
the field distributions. We make the ansatz
$\int\!dh~h^k\Phi(h)=\order(\epsilon^k)$ and
$\int\!dh~h^k\Psi(h)=\order(\epsilon^k)$, with $|\epsilon|\ll 1$,
so that also \begin{eqnarray}
  m&=&
     \beta\int\!\mathrm{d}x~\Phi(x)x+\beta\int\!\mathrm{d}y~\Psi(y)y+\order(\epsilon^3)
\\
  q&=&
      \beta^{2}\int\!\mathrm{d}x\mathrm{d}y~\Phi(x)\Psi(y)(x+y)^{2}+\order(\epsilon^3)
\end{eqnarray}
Bifurcations in which the lowest non-zero order is $\epsilon$ thus
describe transitions towards a ferromagnetic state (P$\to$F), and
those where the lowest non-zero order is $\epsilon^2$ transitions
towards a spin-glass state (P$\to$SG).
 The P$\to$F transitions therefore follow
upon expanding the first order moments of
(\ref{eq:PHI},\ref{eq:PSI}), putting $\overline{x}=\int\!\rmd
x~x\Phi(x)$ and $\overline{y}=\int\!\rmd y~y\Psi(y)$ and using
$A(J,z)=z\tanh(\beta J) +\order(z^3)$:
\begin{eqnarray}
 \overline{x}
  &=&
\tanh(\beta J_0) \overline{x}+ c\bra \tanh(\frac{\beta
J}{c})\ket_J (\overline{x}+\overline{y})
 +\order(\epsilon^3)
\\
\overline{y}
 &=&
\tanh(\beta J_0)\overline{y}+ c \tanh(\beta
J_0)\bra\tanh(\frac{\beta J}{c})\ket_J(\overline{x}+\overline{y})
+\order(\epsilon^3)
\end{eqnarray}
From this one deduces that a continuous $m\neq 0$ bifurcation
occurs at
\begin{eqnarray}
{\rm P}\to {\rm F:} &~~~~~~&
  1= c~\bra\tanh(\frac{\beta J}{c})\ket_{J}~e^{2\beta J_{0}}
  \label{eq:PtoF}
\end{eqnarray}
Similarly we find the P$\to$SG transitions upon expanding the
second order moments of (\ref{eq:PHI},\ref{eq:PSI}), assuming
$\overline{x}=\overline{y}=0$ and putting
$\overline{x^2}=\int\!\rmd x~x^2\Phi(x)$ and
$\overline{y^2}=\int\!\rmd y~y^2\Psi(y)$:
\begin{eqnarray}
 \overline{x^2}
  &=& \tanh^2(\beta J_0) \overline{x^2}+
   c \bra \tanh^2(\frac{\beta J}{c})\ket_J (\overline{x^2}+\overline{y^2}) +\order(\epsilon^3)
\\
\overline{y^2}
 &=&\tanh^2(\beta J_0)\overline{y^2}+c\tanh^2(\beta J_0)
 \bra \tanh^2(\frac{\beta J}{c})\ket_J
(\overline{x^2}+\overline{y^2}) +\order(\epsilon^3)
\end{eqnarray}
From this one deduces that a continuous bifurcation of $q>0$
(while $m=0$)  occurs at
\begin{eqnarray}
{\rm P}\to {\rm SG:} &~~~~~~&
  1= c~\bra\tanh^{2}(\frac{\beta J}{c})\ket_{J}~\cosh(2\beta J_{0})
   \label{eq:PtoSG}
\end{eqnarray}
For $\beta=0$ the right-hand sides of
(\ref{eq:PtoF},\ref{eq:PtoSG}) are both zero, so the actual
(second order) physical transition will occur at the highest
temperature for which either of the two right-hand sides has
increased to unity.

In the next two sections we will work out our theory for two
particular simple choices for the Poissonian long-range bond
distribution $p(J)$.  Numerical evidence suggests that first order
transitions away from the paramagnetic state are absent.

\section{Small world systems with
non-disordered bonds}

We first choose $p(J_{ij})=\delta[J_{ij}-J]$ with $J\geq 0$, i.e.
uniform sparse long-range ferromagnetic bonds (for $J\leq 0$ there
can obviously never be any transition).

\subsection{The phase diagram}

The second order transition lines   (\ref{eq:PtoF}) and
(\ref{eq:PtoSG}) now reduce to
\begin{eqnarray}
{\rm P}\to {\rm F:} &~~~~~~&
  1= c~\tanh(\frac{\beta J}{c})~e^{2\beta J_{0}}
\\
{\rm P}\to {\rm SG:} &~~~~~~&
  1= c~\tanh^{2}(\frac{\beta J}{c})~\cosh(2\beta J_{0})
\end{eqnarray}
Since $\cosh(2\beta J_0)\leq e^{2\beta J_0}$ and $\tanh^2(\beta
J/c)\leq \tanh(\beta J/c)$ for $J\geq 0$,  the P$\to$SG
instability cannot occur as it will always be preceded by the
P$\to$F one. As expected we find only the P and F phases. The
equation from which to solve the transition temperature $T_{\rm
F}$ can be rewritten as
\be
  \beta J=\frac{1}{2}c~\log\left[\frac{c+e^{-2\beta J_{0}}}{c-e^{-2\beta
  J_{0}}}\right]
 \label{eq:ferroPF}
\ee In the limit $c\to\infty$ we find condition (\ref{eq:ferroPF})
reducing to $\beta J=e^{-2\beta J_{0}}$. This expression has been
found earlier in the context of $1+\infty$ dimensional attractor
neural networks \cite{skantzos-coolen00}.

\begin{figure}[t] \vspace*{9mm} \hspace*{45mm}
\setlength{\unitlength}{0.65mm}
\begin{picture}(100,100)
\put(5,5){\epsfysize=100\unitlength\epsfbox{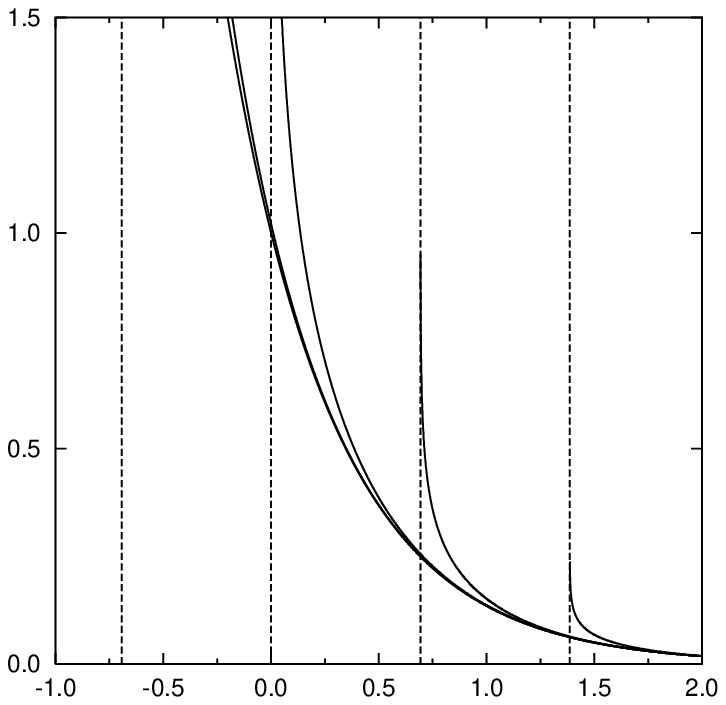}}
\put(60,-6){\here{$\beta J_0$}} \put(1,55){\here{$\beta J$}}

 \put(91,108){\here{\small $c=\frac{1}{16}$}}
 \put(69,108){\here{\small $c=\frac{1}{4}$}}
 \put(49,108){\here{\small $c=1$}}

 \put(26,30){\here{\large P}}\put(100,70){\here{\large F}}
\end{picture}
\vspace*{9mm} \caption{Phase diagram in the ($\beta J_0,\beta J)$
plane of the `small world' magnet (i.e. uniform sparse long-range
bonds), defined by $H=-J_{0}\sum_{i}\sigma_{i}\sigma_{i+1}
             -\frac{J}{c}\sum_{i<j}c_{ij}\sigma_{i}\sigma_{j}$.
Solid lines: P$\to$F transition lines for
$c=\frac{1}{16},\frac{1}{4},1,4,16$ (from right to left). Dashed
vertical lines: asymptotic values $\beta J_0=\log(1/\sqrt{c})$ at
which the critical values of  $\beta J$ diverge (and  below which
no order is possible), for $c=\frac{1}{16},\frac{1}{4},1,4$ (from
right to left).  This diagram demonstrates the so called `small
world effect'. Along the axis $\beta J=0$ we have a
one-dimensional model, where there is no phase transition at
finite temperature. Along the $\beta J_0=0$ axis we have the
random finitely connected graph, where a P$\to$F transition is
known to exist at $\beta J =\frac{1}{2}c\log[(c+1)/(c-1)]$,
provided $c>1$ (i.e. above the percolation threshold). When the
two structures are combined we always find a phase transition at
finite temperature, for any value of $c$ (however small). }
\label{fig:uniform_diagram}
\end{figure}

We note in  (\ref{eq:ferroPF}) that ferromagnetic order requires
$\beta J_0>\log(1/\sqrt{c})$; this inequality always holds when
$c>1$, and puts a finite upper bound on $T_{\rm F}$ when $c<1$.
Given $\beta J_0>\log(1/\sqrt{c})$ and $J_0>0$, however, there
will now always be a solution $T_{\rm F}>0$.  The simple equation
(\ref{eq:ferroPF}) thus reveals  the `small world' effect in
statistical mechanical terms: for any nonzero average Poissonian
connectivity $c$ (including $c<1$) and all non-zero bond strengths
$\{J_0,J\}$ (however small) there exists a {\em finite} transition
temperature $T_{\rm F} (J_0,J,c)$ where ferromagnetism sets in.
Since the pure Poissonian graph (i.e. $J_0=0$) would exhibit a
transition only when $c>1$, above the percolation threshold, this
result is not obvious. One can apparently induce an overall
non-zero magnetization in the one-dimensional ferromagnetic ring
by adding a very small number of long range connections.
Similarly, the one-dimensional ring leads to the emergence of a
non zero magnetization even when superimposed on random graphs
with $c<1$, below the percolation threshold.

We have drawn the line (\ref{eq:ferroPF}) in a $(\beta J_0,\beta
J)$ phase diagram, for different values of the average Poissonian
connectivity $c$, in figure \ref{fig:uniform_diagram}. Lowering
the temperature from $T=\infty$ down to $T=0$ corresponds in this
diagram to moving along a straight line, with slope given by
$J/J_0$, away from the origin $(\beta J_0,\beta J)=(0,0)$.

\subsection{Effective field distributions}

\begin{figure}[t]
\setlength{\unitlength}{0.6mm}\hspace*{12mm}
\begin{picture}(200,160)
\put(0,80){\includegraphics[height=77\unitlength,width=80\unitlength]{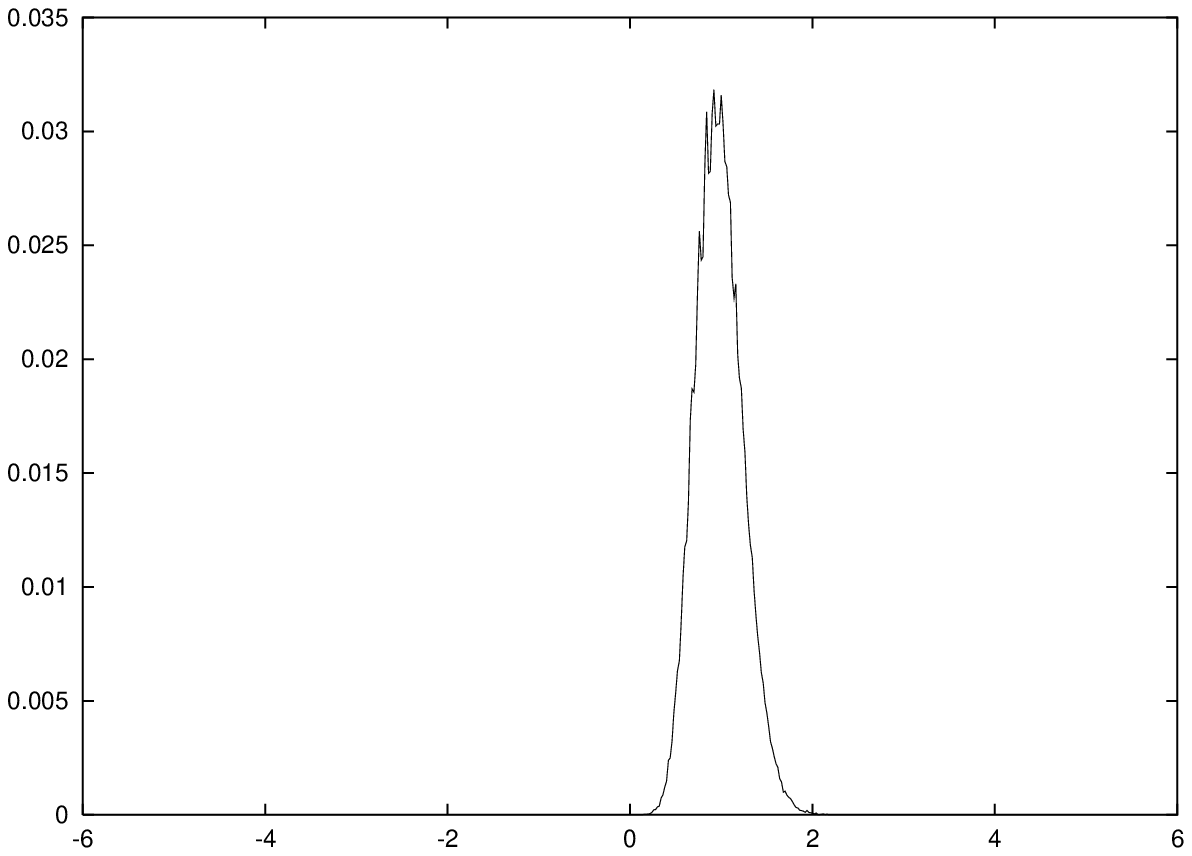}}
\put(80,80){\includegraphics[height=77\unitlength,width=80\unitlength]{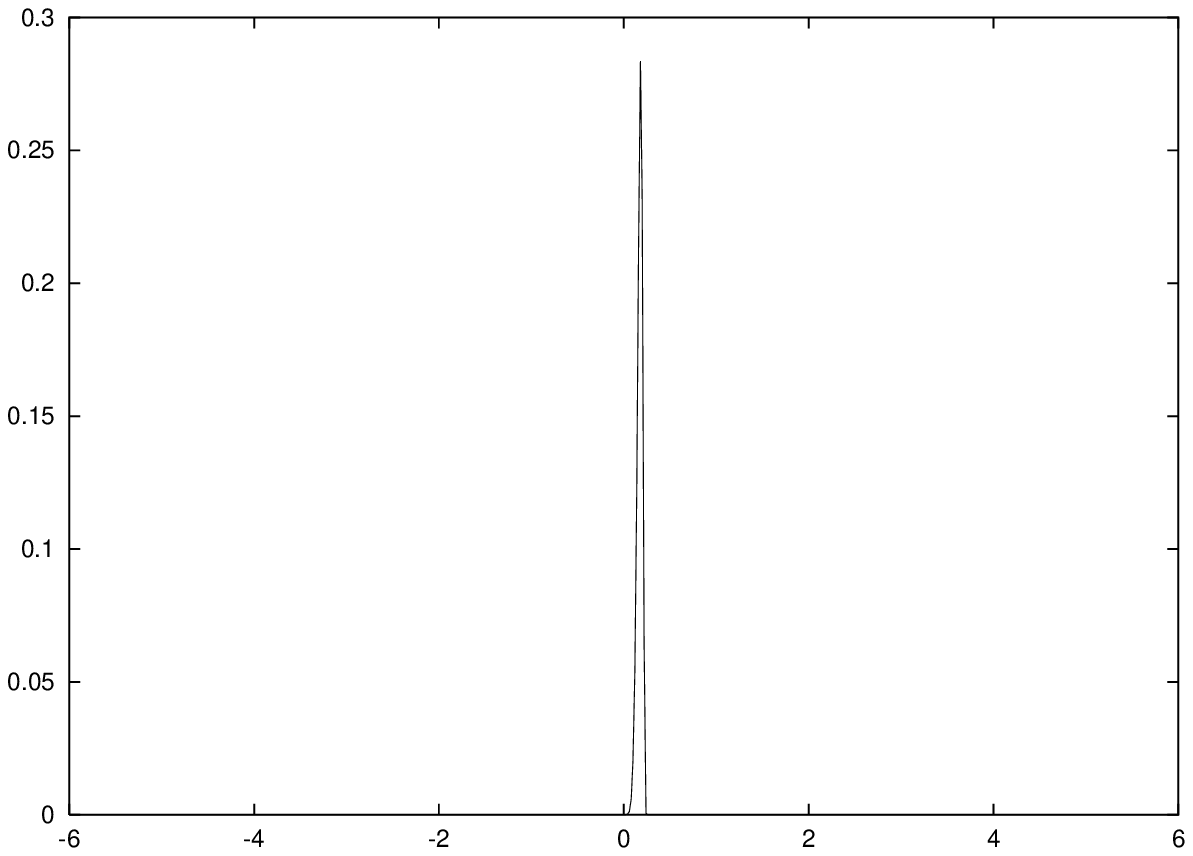}}
\put(160,80){\includegraphics[height=77\unitlength,width=80\unitlength]{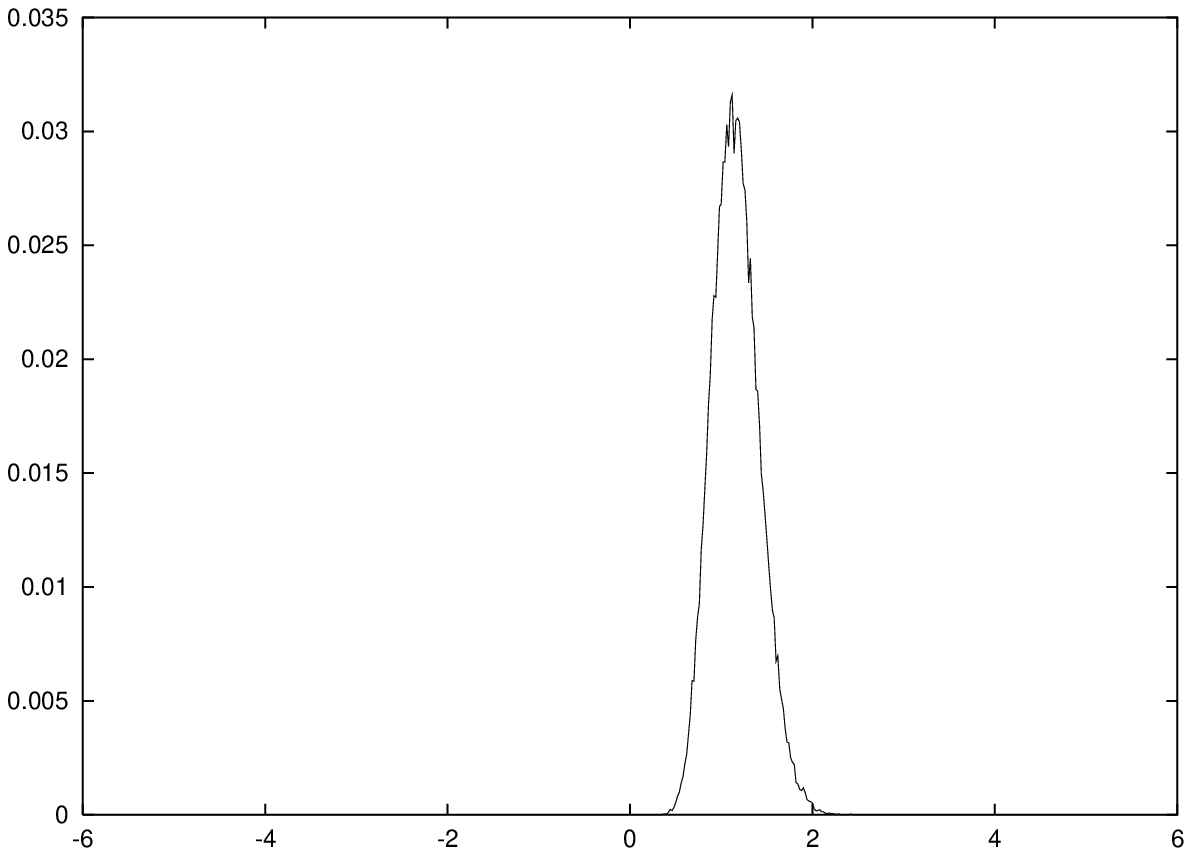}}

\put(0,0){\includegraphics[height=77\unitlength,width=80\unitlength]{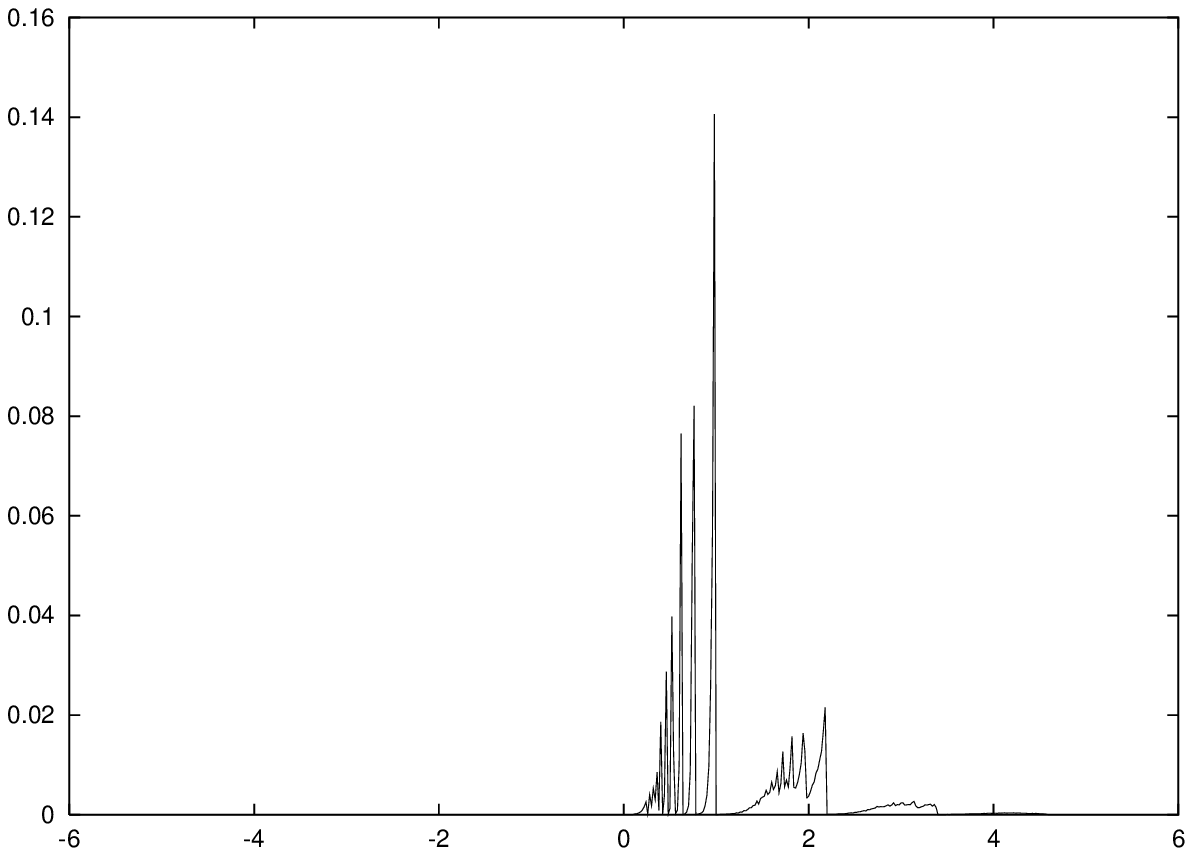}}
\put(80,0){\includegraphics[height=77\unitlength,width=80\unitlength]{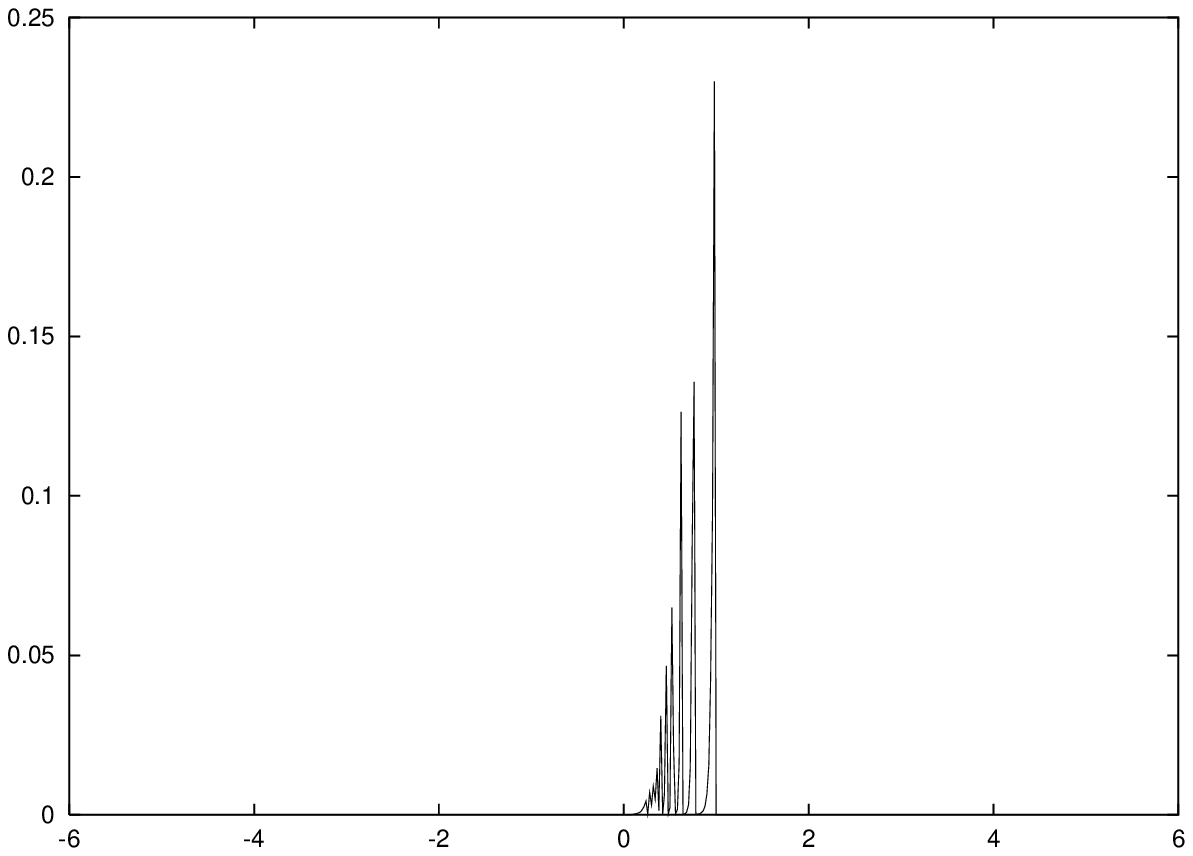}}
\put(160,0){\includegraphics[height=77\unitlength,width=80\unitlength]{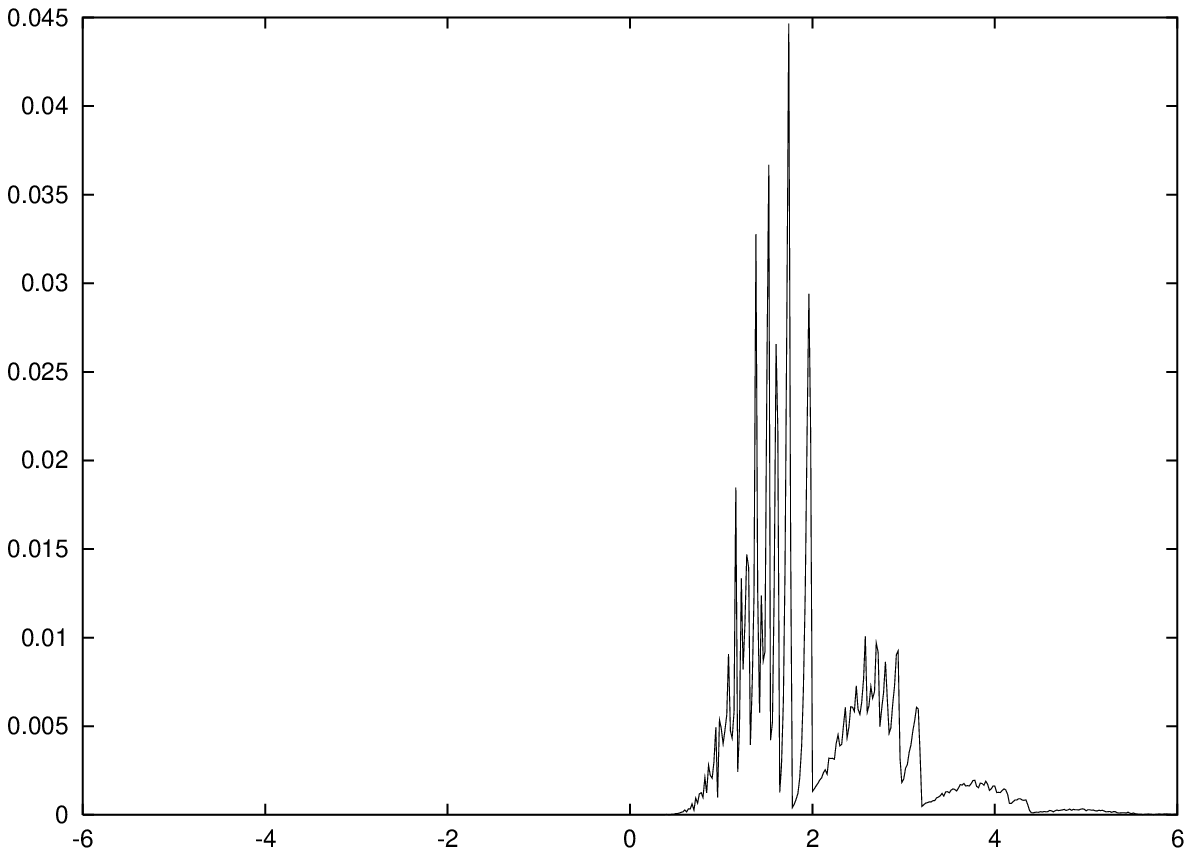}}

\put(42,-8){\here{\small $h$}} \put(122,-8){\here{\small
$h$}}\put(202,-8){\here{\small $h$}}

\put(20,65){\here{$\Phi(h)$}} \put(100,65){\here{$\Psi(h)$}}
\put(180,65){\here{$W(h)$}}

\put(20,145){\here{$\Phi(h)$}} \put(100,145){\here{$\Psi(h)$}}
\put(180,145){\here{$W(h)$}}

\end{picture}
\vspace*{8mm}
 \caption{Effective field distributions for the `small world'
ferromagnet (i.e. uniform sparse long-range bonds), defined by
$H=-J_{0}\sum_{i}\sigma_{i}\sigma_{i+1}
             -\frac{J}{c}\sum_{i<j}c_{ij}\sigma_{i}\sigma_{j}$,
            as obtained by numerical solution of the order
parameter equations (\ref{eq:PHI},\ref{eq:PSI},\ref{eq:W}) in the
ferromagnetic phase.
 Top row: distributions well above the percolation threshold of the Poissonian graph ($T=1$, $c=10$,
$J_0=\frac{1}{4}$ and $J=1$). Bottom row: distributions for a
system below the percolation threshold and for more dominant chain
bonds ($T=\frac{3}{2}$, $c=\frac{1}{2}$, $J_0=1$ and
$J=\frac{3}{5}$). Note that the vertical scales used in the two
rows are not identical.} \label{fig:pop_uniform}
\end{figure}

To appreciate the physics in the ordered state of the small world
magnet, we also solved numerically the order parameter equations
(\ref{eq:PHI},\ref{eq:PSI},\ref{eq:W}), which here simplify to
\begin{eqnarray}
\hspace*{-15mm}
 \Phi(x)
  &=& \int\!\rmd x^\prime~ \Phi(x^\prime)
 \sum_{k\geq 0}\frac{ e^{-c} c^k}{k!}
 \int\!\prod_{\ell=1}^k\left[ \rmd h_\ell
W(h_\ell)\right]
     \delta[
     x- A(J_0,x^\prime)- \sum_{\ell=1}^k
A(\frac{J}{c},h_\ell) ] \label{eq:Phi_uniform}
\\
\hspace*{-15mm} \Psi(y)
 &=&
 \int\!\rmd y^\prime ~\Psi(y^\prime)
\sum_{k\geq 0}\frac{ e^{-c} c^k}{k!}
 \int\!\prod_{\ell=1}^k\left[ \rmd h_\ell
W(h_\ell) \right] \delta[y-A(J_0,y^\prime\!+\sum_{\ell=1}^k
A(\frac{J}{c},h_\ell))] \label{eq:Psi_uniform}
\\
\hspace*{-15mm}
  W(h)&=&\int\!\mathrm{d}x\mathrm{d}y~\Phi(x)\Psi(y)\delta[h-x-y]
  \label{eq:W_uniform}
\end{eqnarray}
The results are shown in figure \ref{fig:pop_uniform} for two
choices of control parameters  in the ferromagnetic region: for
$T=1$, $c=10$, $J_0=\frac{1}{4}$ and $J=1$  (i.e. above the
percolation threshold of the Possionian graph) and for
$T=\frac{3}{2}$, $c=\frac{1}{2}$, $J_0=1$ and $J=\frac{3}{5}$
(i.e. below the percolation threshold). In the first case the
effective field distributions are found to be smooth. For $c<1$,
however, the distributions are seen to be highly irregular.

\subsection{Comparison with simulations}

\begin{figure}[t]
\vspace*{5mm}

\setlength{\unitlength}{0.65mm}\hspace*{23mm}
\begin{picture}(200,80)
\put(10,0){\includegraphics[height=85\unitlength,width=90\unitlength]{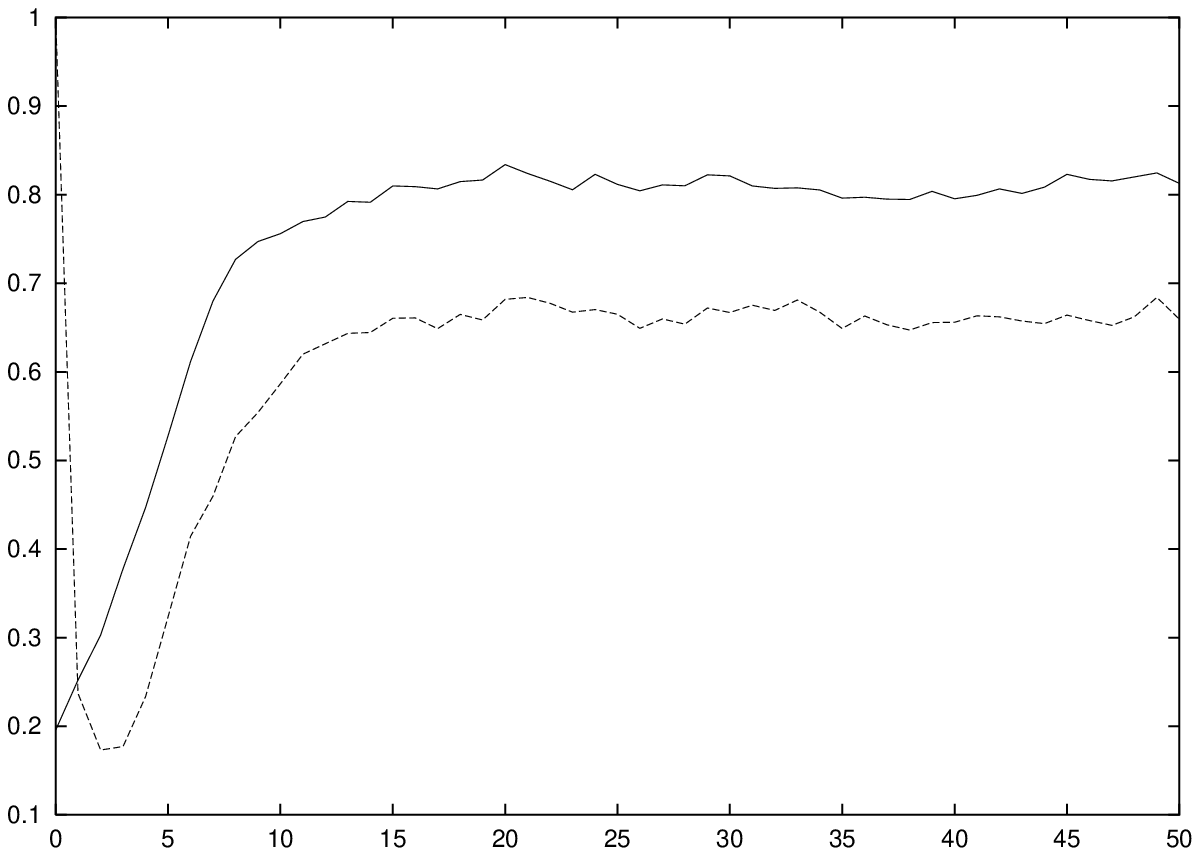}}
\put(109,0){\includegraphics[height=85\unitlength,width=90\unitlength]{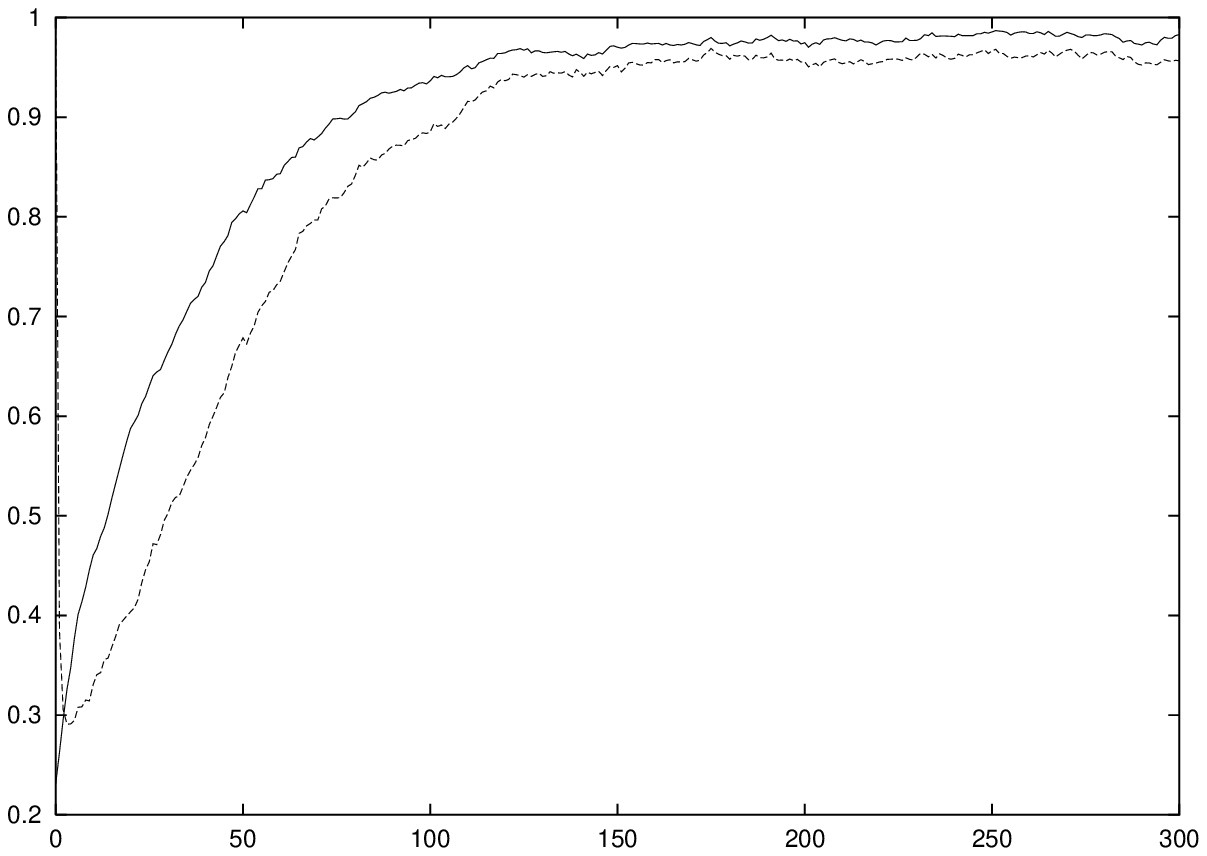}}

\put(58,-8){\here{\small $t$ (iter/spin)}}
\put(157,-8){\here{\small $t$ (iter/spin)}}

\put(-3,46){\here{$m,~q$}}
\end{picture}
\vspace*{10mm}
 \caption{Observables $m$ (solid lines) and $q$ (dashed lines)
  as measured during numerical simulations of the `small world' magnet,
 for $N=10,\!000$.
 Left: $T=1$, $c=10$,
$J_0=\frac{1}{4}$ and $J=1$. Here the theory predicts the
equilibrium values $m=0.80$ and $q=0.65$. Right: $T=\frac{2}{3}$,
$c=\frac{1}{2}$, $J_0=1$ and $J=\frac{3}{5}$. Here the theory
predicts the equilibrium values $m\simeq 0.98$ and $q\simeq 0.96$.
In both cases equilibration is achieved relatively fast, and the
agreement between theory and simulations is very good.}
\label{fig:sim_uniform}
\end{figure}

To complete our analysis for the small world magnet we have
compared the predicted values for $m$ and $q$ of our solution as
obtained by numerical solution of
(\ref{eq:Phi_uniform},\ref{eq:Psi_uniform},\ref{eq:W_uniform}),
followed by evaluation of (\ref{eq:q_and_m}), with the result of
numerical simulation of the stochastic microscopic dynamics,
 taken to be of the conventional Glauber type and with $N=10,\!000$. In the
present (non-frustrated) small world magnet equilibration is found
to be relatively easy to achieve, as a result of which the
agreement between theory and experiment is found to be very good.
Examples are shown in figure \ref{fig:sim_uniform}, for parameter
values identical to those used earlier to produce the field
distributions in figure \ref{fig:pop_uniform}.

\section{Small world spin glasses}

We finally work out our theoretical predictions for a `small
world' spin-glass, where the sparse long range bonds are truly
random, and distributed according to
$p(J_{ij})=p\delta(J_{ij}-J)+(1-p)\delta(J_{ij}+J)$. Without loss
of generality we may take $J\geq 0$.

\subsection{The phase diagram}

\begin{figure}[t] \vspace*{9mm} \hspace*{45mm}
\setlength{\unitlength}{0.65mm}
\begin{picture}(100,100)
\put(5,5){\epsfysize=100\unitlength\epsfbox{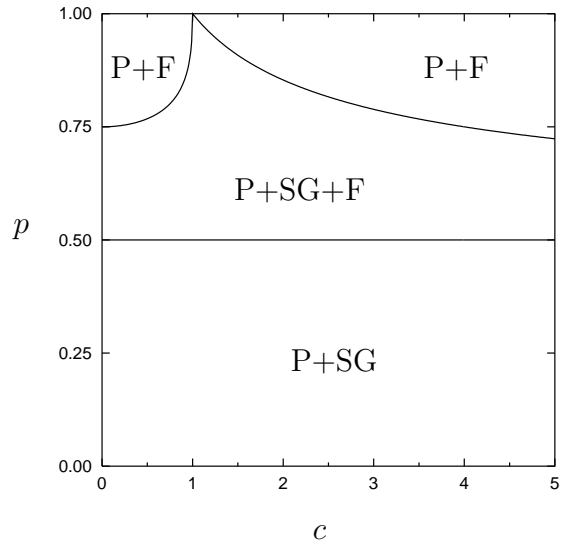}}
\put(62,-5){\here{$c$}} \put(1,57){\here{$p$}}

 \put(65,30){\here{P+SG}}\put(58,65){\here{P+SG+F}}
 \put(26,90){\here{P+F}} \put(90,90){\here{P+F}}
\end{picture}
\vspace*{7mm} \caption{The possible phases of the `small world'
spin-glass  for the different allowed combinations of the
dimensionless control parameters $c\geq 0$ (the average Poissonian
connectivity) and $p\in[0,1]$ (the probability of $J_{ij}=+J$ in
the Poissonian bonds). } \label{fig:pc_regimes}
\end{figure}

The second order transition lines   (\ref{eq:PtoF}) and
(\ref{eq:PtoSG}) now take the form
\begin{eqnarray}
{\rm P}\to {\rm F:} &~~~~~~&
  1= c(2p-1)\tanh(\frac{\beta J}{c})~e^{2\beta J_{0}}
\\
{\rm P}\to {\rm SG:} &~~~~~~&
  1= c~\tanh^{2}(\frac{\beta J}{c})~\cosh(2\beta J_{0})
\end{eqnarray}
Now both transitions may generally occur, dependent on the values
of $p$ and $c$, unless $p\leq\frac{1}{2}$ where only the P$\to$SG
transition is possible. For $c\to\infty$ we see that only the
P$\to$F transition will remain. We will transform the above
equations into the dimensionless variables $x=J_0/J$ and $y=T/J$,
which gives for the P$\to$F transition
\begin{eqnarray}
{\rm P}\to {\rm F:} &~~~~~~& x_{\rm F}=\frac{1}{2}y
\log\left[(2p-1)c\tanh(1/yc)\right]^{-1}~~~~~~(p\geq
\frac{1}{2}~{\rm only})
\\
&& \hspace*{-9mm} y\to 0: ~~~~~ x_{\rm
F}=\frac{1}{2}y\log\left[(2p-1)c\right]^{-1}+\ldots\nonumber \\ &&
\hspace*{-9mm} y\to \infty: ~~~ x_{\rm
F}=\frac{1}{2}y\log\left[y/(2p-1)\right]+\ldots \nonumber
\end{eqnarray}
The line will intersect the $x=0$ axis at
$y=1/c\log\sqrt{[(2p-1)c+1]/[(2p-1)c-1]}$, provided $c<1/(2p-1)$.
The P$\to$SG transition line now takes the form
\begin{eqnarray}
{\rm P}\to {\rm SG:} &~~~~~&
 x_{\rm SG}=\frac{1}{2}y
\log\left[\frac{1+\sqrt{1-c^2\tanh^4(1/yc)}}{c\tanh^2(1/yc)}\right]
\\
&& \hspace*{-9mm} y\to 0: ~~~~~ x_{\rm
SG}=\frac{1}{2}y\log\left[(1+\sqrt{1-c^2})/c\right]+\ldots~~~~(c\leq
1~{\rm only})\nonumber
\\ && \hspace*{-9mm} y\to \infty: ~~~ x_{\rm
SG}=\frac{1}{2}y\log\left[2cy^2\right]+\ldots \nonumber
\end{eqnarray}
Clearly, the physics for $c>1$ will again be significantly
different from that found for $c\leq 1$. If $c\leq 1$ one will
have a P$\to$SG transition for any $y>0$, whereas for $c>1$ it
will only occur when $y>2/c\log[(\sqrt{c}+1)/(\sqrt{c}-1)]$. The
two transition lines can meet at a triple point, which is found to
be the solution of
\be
\hspace*{-15mm} {\rm
triple~point}~(x^\star,y^\star):~~~\left\{\begin{array}{l}
\tanh(1/y^\star c)=(2p-1)\left[1+\sqrt{1-c^2\tanh^4(1/y^\star
c)}\right]\\
 x^\star=\frac{1}{2}y^\star
\log\left[(2p-1)c\tanh(1/y^\star c)\right]^{-1}
\end{array}\right.
 \label{eq:triple} \ee Graphical inspection of the
equation $\tau=(2p-1)[1+\sqrt{1-c^2\tau^4}]$ shows that the
necessary and sufficient  existence condition for the triple point
are:
\be
c\leq 1:~~~~2p-1\leq \frac{1}{1+\sqrt{1-c^2}},~~~~~~~~c\geq
1:~~~~2p-1\leq \frac{1}{\sqrt{c}} \label{eq:SGphaseconditions} \ee
We may conclude, since $\lim_{y\to \infty} x_{\rm F}/x_{\rm
SG}=0$, that as long as $p> \frac{1}{2}$ there will be a P$\to$F
transition, but that a P$\to$SG transition exists if and only if
(\ref{eq:SGphaseconditions}) is satisfied (if not, the P$\to$F
transition will always happen first). Thus, the possible phases
(dependent on the remaining energy-related control parameters
$T,J,J_0$) are \bd
 p\leq \frac{1}{2}\!:~~\{{\rm P},{\rm SG}\}
 ~~~~~~~~~
 \frac{1}{2}< p<p_c\!:~~ \{{\rm P},{\rm SG},{\rm F}\}~~~~~~~
p\geq p_c\!:~~ \{{\rm P},{\rm F}\} \ed where \bd
 p_{c\leq
1}=\frac{1}{2}+\frac{1}{2(1+\sqrt{1-c^2})} ~~~~~~~~ p_{c\geq
1}=\frac{1}{2}+ \frac{1}{2\sqrt{c}} \ed This is summarized in
figure \ref{fig:pc_regimes}.
 We are not yet able to
determine the F$\to$SG transition (if both F and SG phases exist)
analytically, since this would require us to solve our equations
also below the P$\to$F and/or P$\to$SG transition temperatures.
However we may put forward the conjecture (which seems reasonable
on the basis of our experience with more conventional disordered
spin models, e.g. \cite{SK,RSB,WatkinSherr}),
 that, especially upon taking
RSB into account (if needed),  there will be no change of phase
type after the onset of order as the temperature is lowered from
$T=\infty$ to $T=0$. This conjecture would predict the elusive
F$\to$SG transition to be the line segment in the $(x,y)$ plane
going from $(x^\star,0)$ to $(x^\star,y^\star)$, where the latter
is the triple point (\ref{eq:triple}).

\begin{figure}[t]
\setlength{\unitlength}{0.6mm}\hspace*{12mm}
\begin{picture}(200,170)

 \put(16,140){P}\put(96,140){P}\put(176,140){P}
 \put(16,60){P}\put(96,60){P}\put(176,60){P}

\put(20,95){SG}\put(140,100){F}\put(97,95){SG} \put(210,100){F}

\put(50,20){SG}
 \put(140,20){F}\put(105,11){SG}\put(210,20){F}

\put(0,80){\includegraphics[height=77\unitlength,width=110\unitlength]{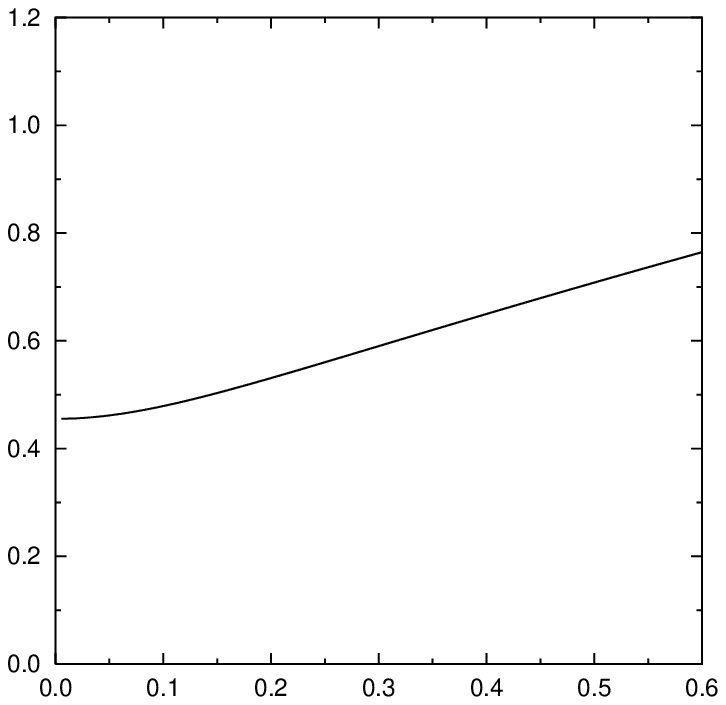}}
\put(80,80){\includegraphics[height=77\unitlength,width=110\unitlength]{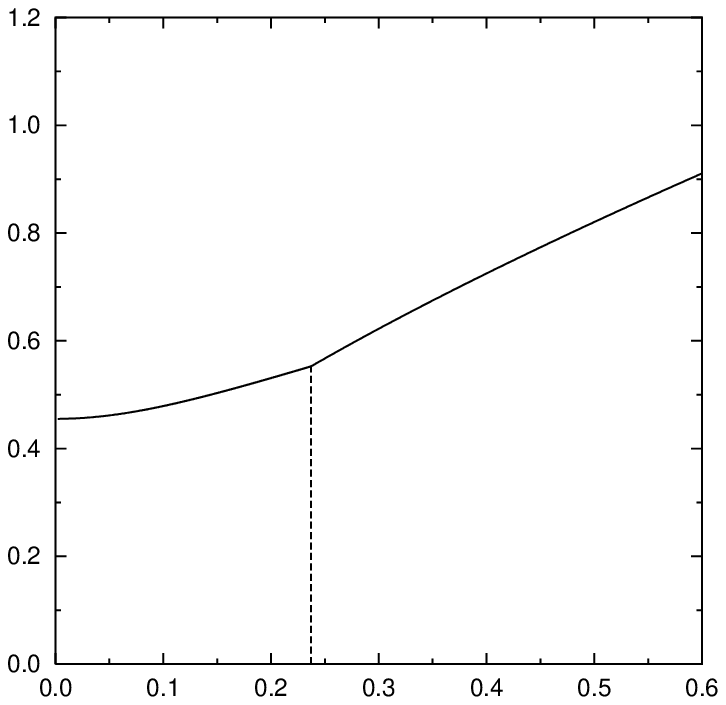}}
\put(160,80){\includegraphics[height=77\unitlength,width=110\unitlength]{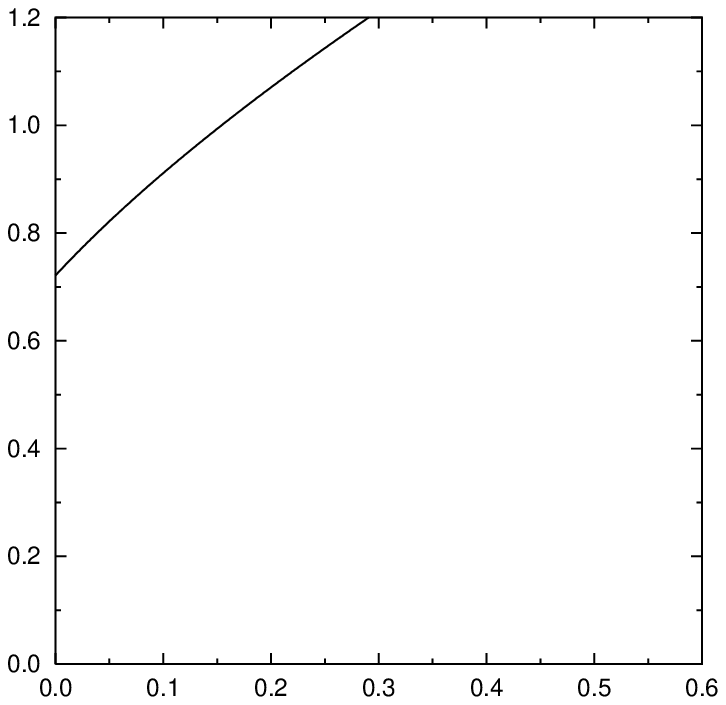}}

\put(2,0){\includegraphics[height=77\unitlength,width=110\unitlength]{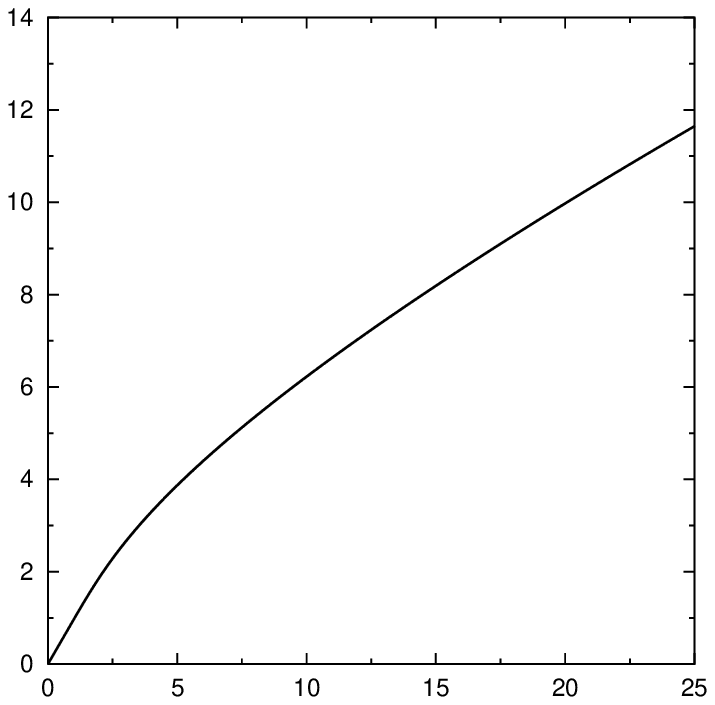}}
\put(82,0){\includegraphics[height=77\unitlength,width=110\unitlength]{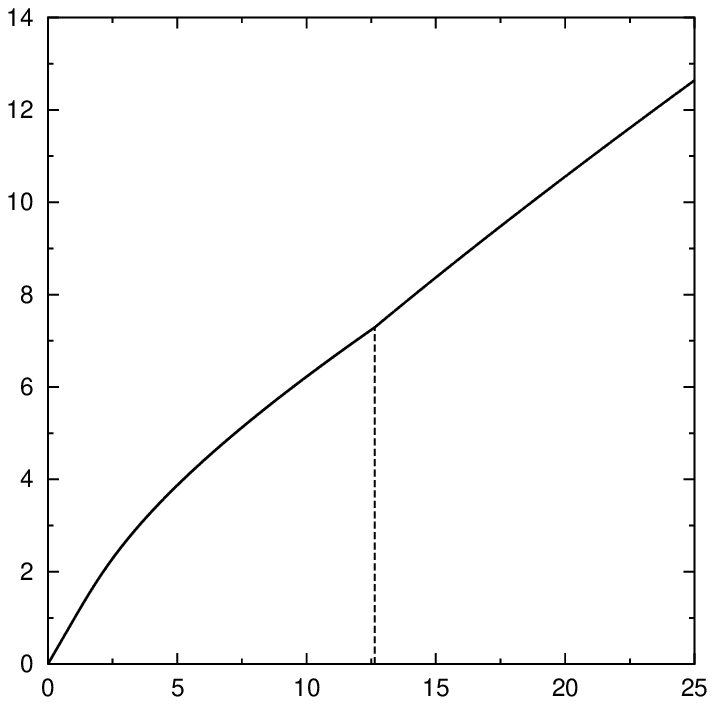}}
\put(162,0){\includegraphics[height=77\unitlength,width=110\unitlength]{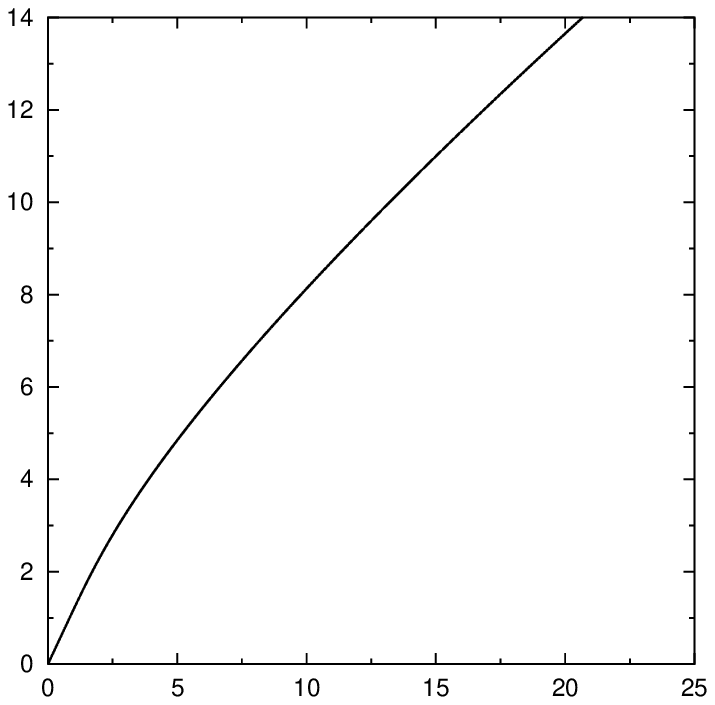}}

\put(42,-9){\here{\small $J_0/J$}} \put(122,-9){\here{\small
$J_0/J$}}\put(202,-9){\here{\small $J_0/J$}}

\put(42,165){\here{$p=1/4$}}
\put(122,165){\here{$p=5/8$}}\put(202,165){\here{$p=7/8$}}

\put(-7,40){\here{\small $T/J$}} \put(-7,120){\here{\small $T/J$}}

\end{picture}
\vspace*{10mm}
 \caption{Phase diagrams in the $(J_0/J,T/J)$ plane for `small world' spin glasses, i.e. ferromagnetic rings with random sparse long-range
 bonds with strengths distributed according to $p(J_{ij})=p\delta[J_{ij}-J]+(1-p)\delta[J_{ij}+J]$. Here $p\in\{\frac{1}{4},\frac{5}{8},\frac{7}{8}\}$.
 The P$\to$F and P$\to$SG transitions (solid lines) and the triple point
 follow from our theory. The F$\to$SG transition (dashed) is
 obtained upon using the location of the triple point in combination
 with the conjecture that on lowering temperature the nature of the ordered phase will
 remain that which emerges at the onset.
 Top row: $c=4$, well above the percolation threshold of the Poissonian
 graph.
 Bottom row: $c=1/4$, well below the percolation threshold. Here neither the ring nor the Poissonian
 graph would have exhibited order, whereas the combination does (the `small world' effect). }
\label{fig:sg_diagrams}
\end{figure}

In figure \ref{fig:sg_diagrams} we show typical examples of the
resulting phase diagrams in the $(J_0/J,T/J)$ plane, probing
systematically the regimes exhibited in figure
\ref{fig:pc_regimes}. It should be noted that, in spite of the
equivalent notation and the resulting temptation to make hasty
comparisons, the parameter $J$ only has a meaning identical to
that in the SK spin-glass \cite{SK}, i.e. measuring the variance
in the bonds, for $p=\frac{1}{2}$. The P$\to$SG instability line
is independent of $p$, in contrast to the P$\to$F one.  We observe
here that, as the fraction $p$ of positive long-range bonds
increases (for fixed connectivity $c$) from $p<\frac{1}{2}$ (where
there can be no ferromagnetic phase) to larger values, the
ferromagnetic phase becomes increasingly important, to the point
where it destroys the spin-glass phase altogether. However, there
is again a clear difference between $c>1$, where even without the
ring (i.e. for $J_0=0$) an ordered state is possible, and $c<1$
(below the percolation threshold of the Poissonian graph) where as
in the case of the small world magnet  it is the `small world'
effect which generates global order.

\subsection{Effective field distributions}

\begin{figure}[t]
\setlength{\unitlength}{0.6mm}\hspace*{12mm}
\begin{picture}(200,160)
\put(0,80){\includegraphics[height=77\unitlength,width=80\unitlength]{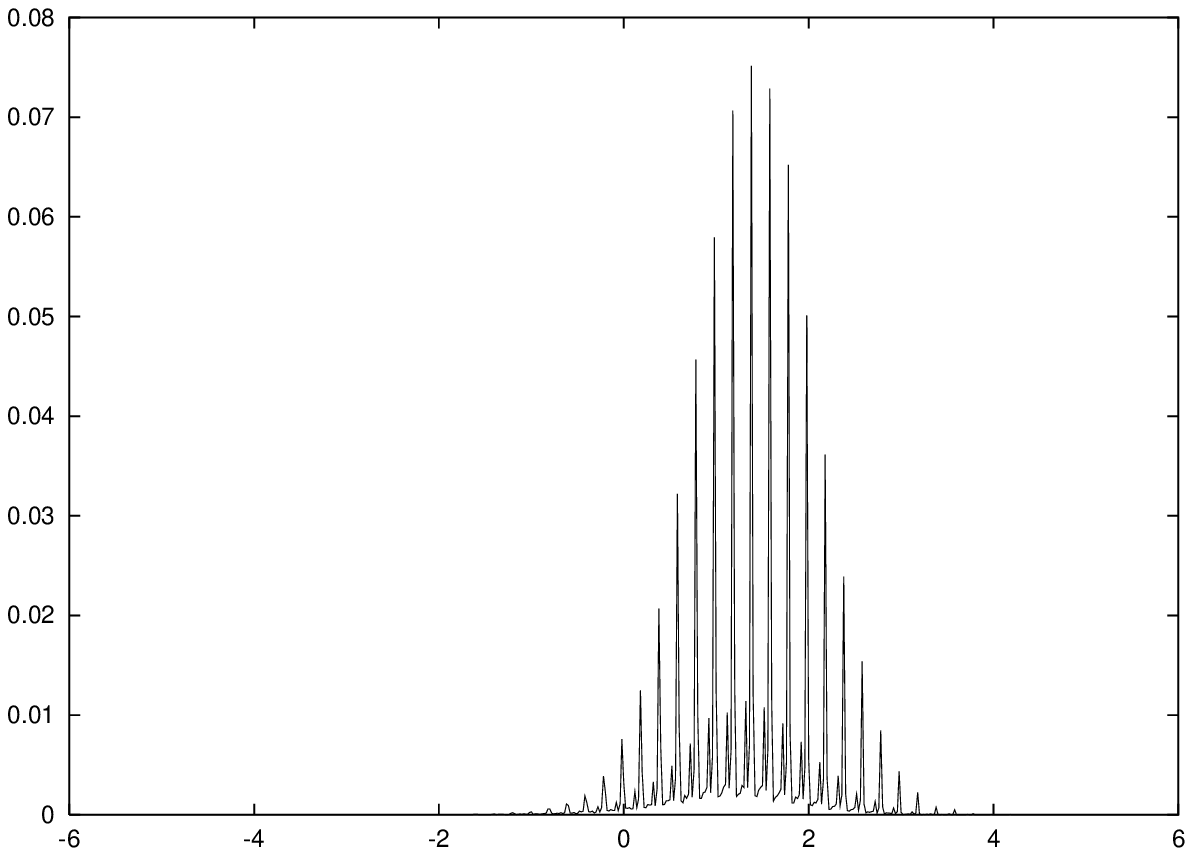}}
\put(80,80){\includegraphics[height=77\unitlength,width=80\unitlength]{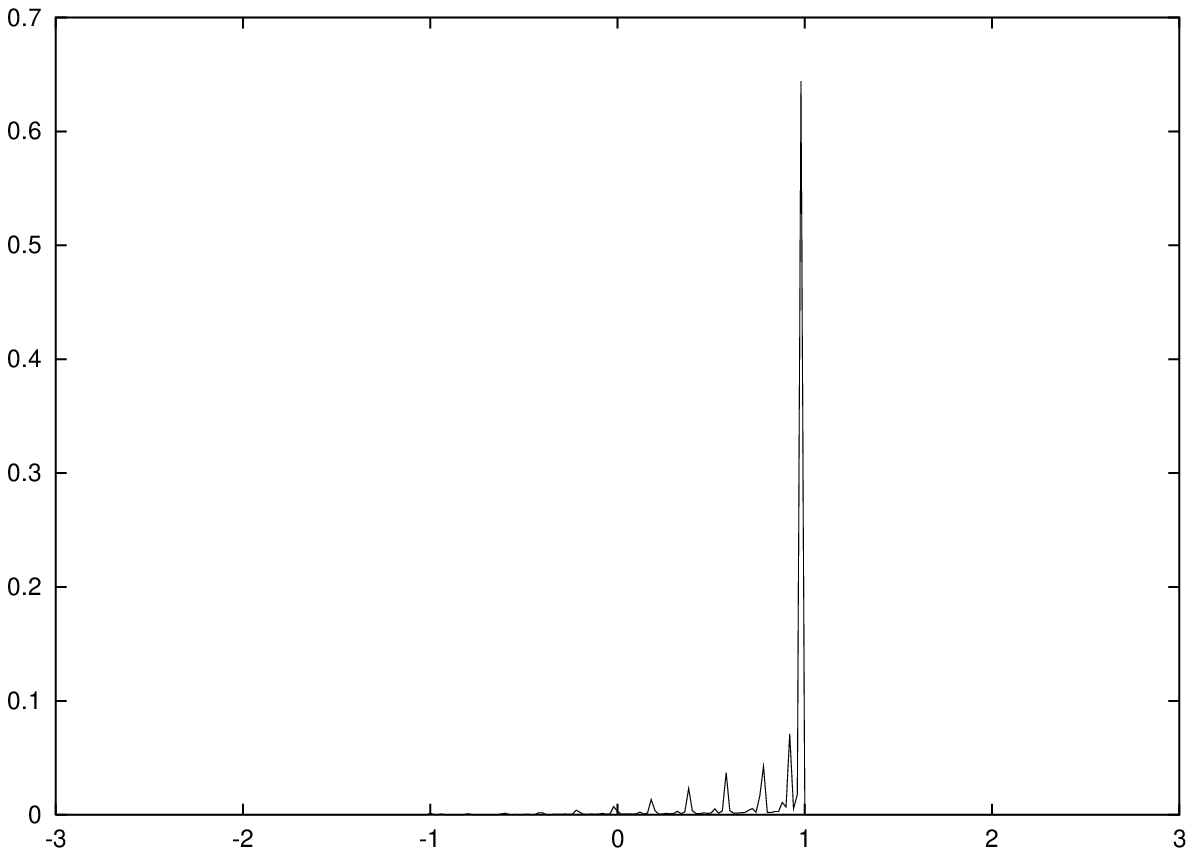}}
\put(160,80){\includegraphics[height=77\unitlength,width=80\unitlength]{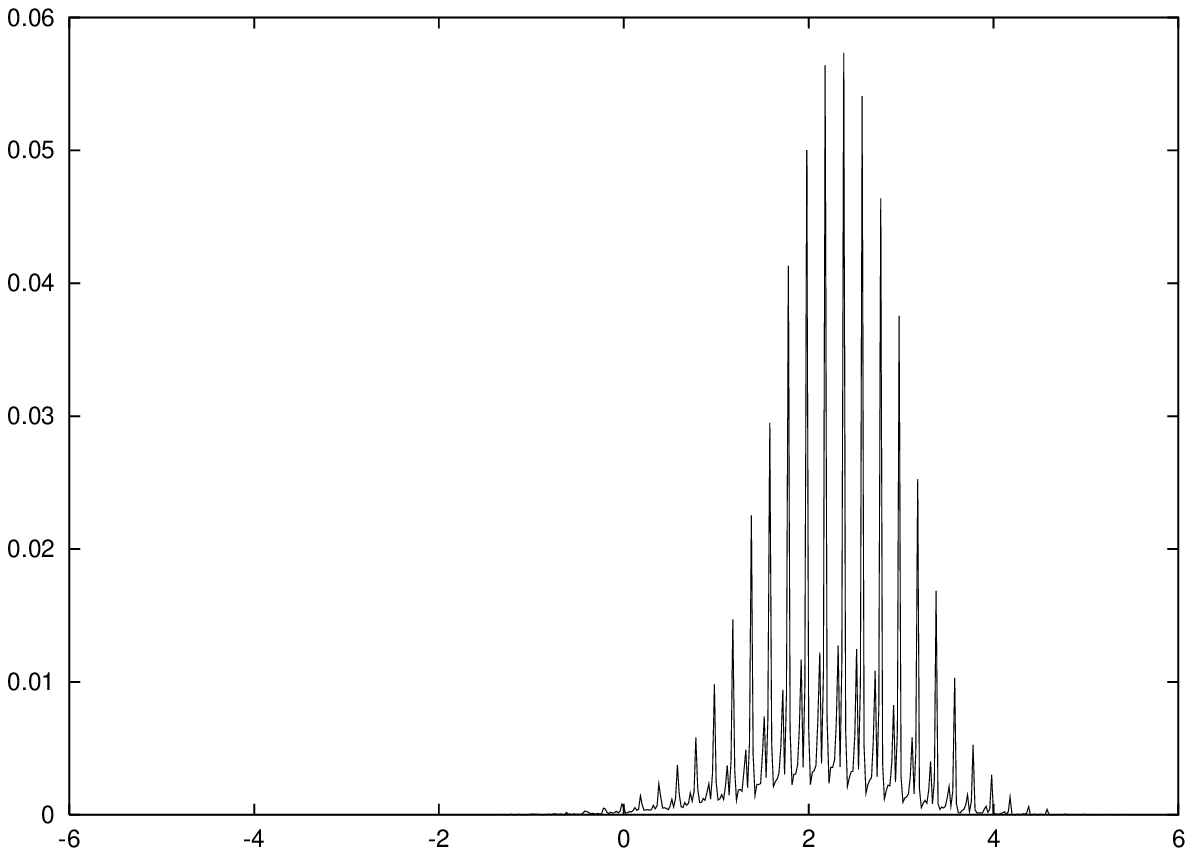}}

\put(0,0){\includegraphics[height=77\unitlength,width=80\unitlength]{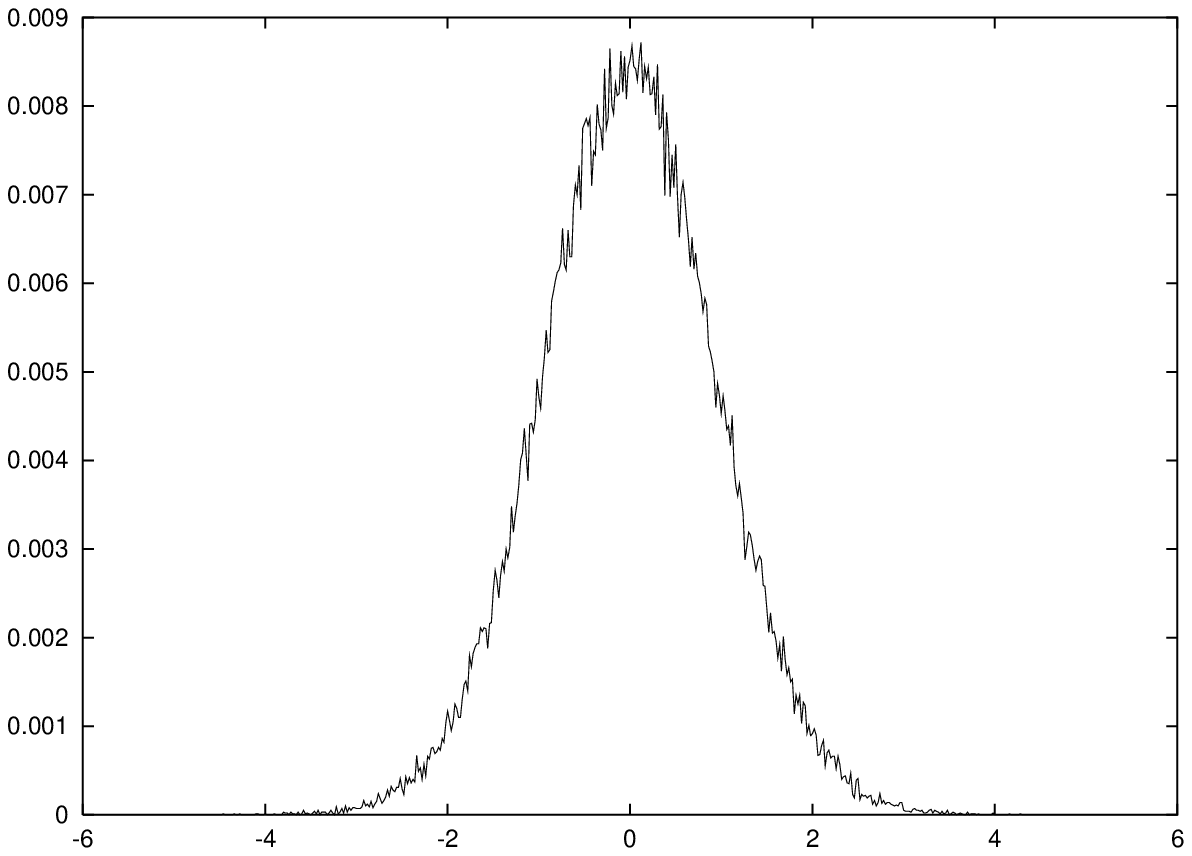}}
\put(80,0){\includegraphics[height=77\unitlength,width=80\unitlength]{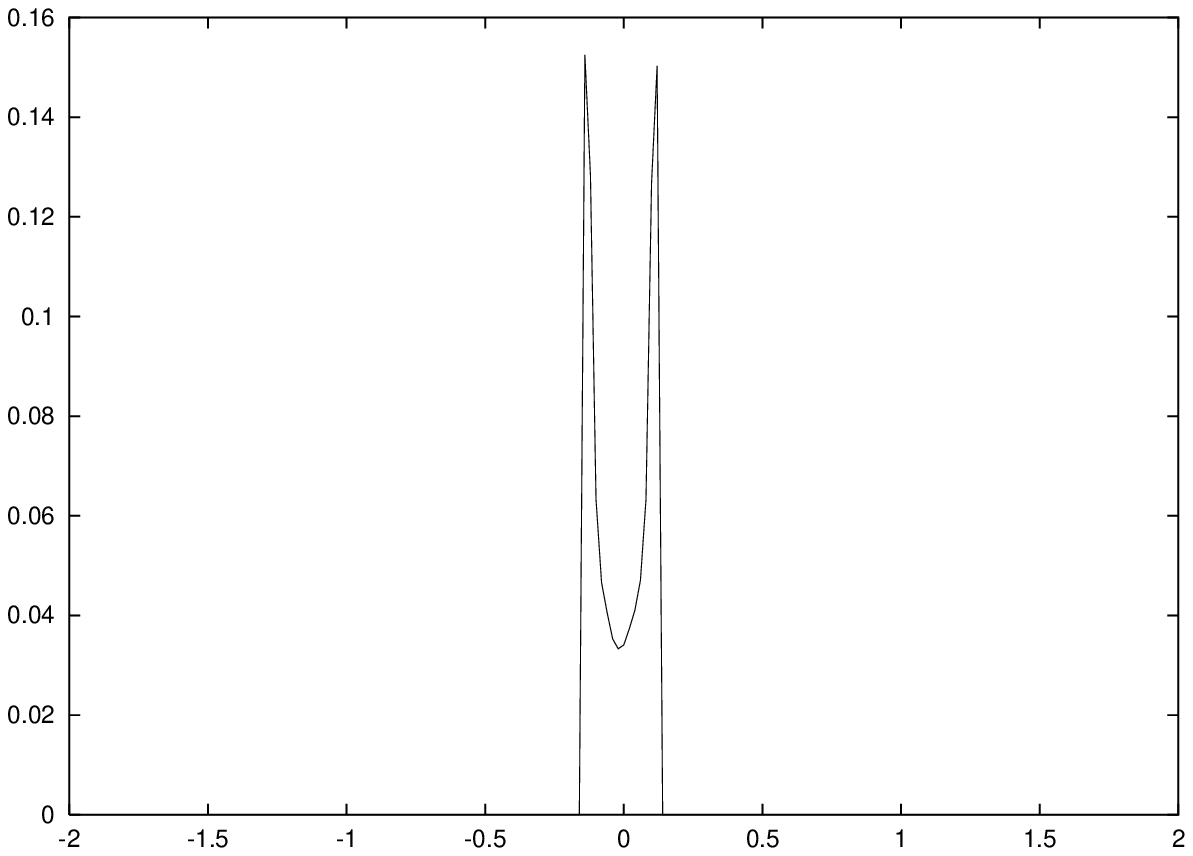}}
\put(160,0){\includegraphics[height=77\unitlength,width=80\unitlength]{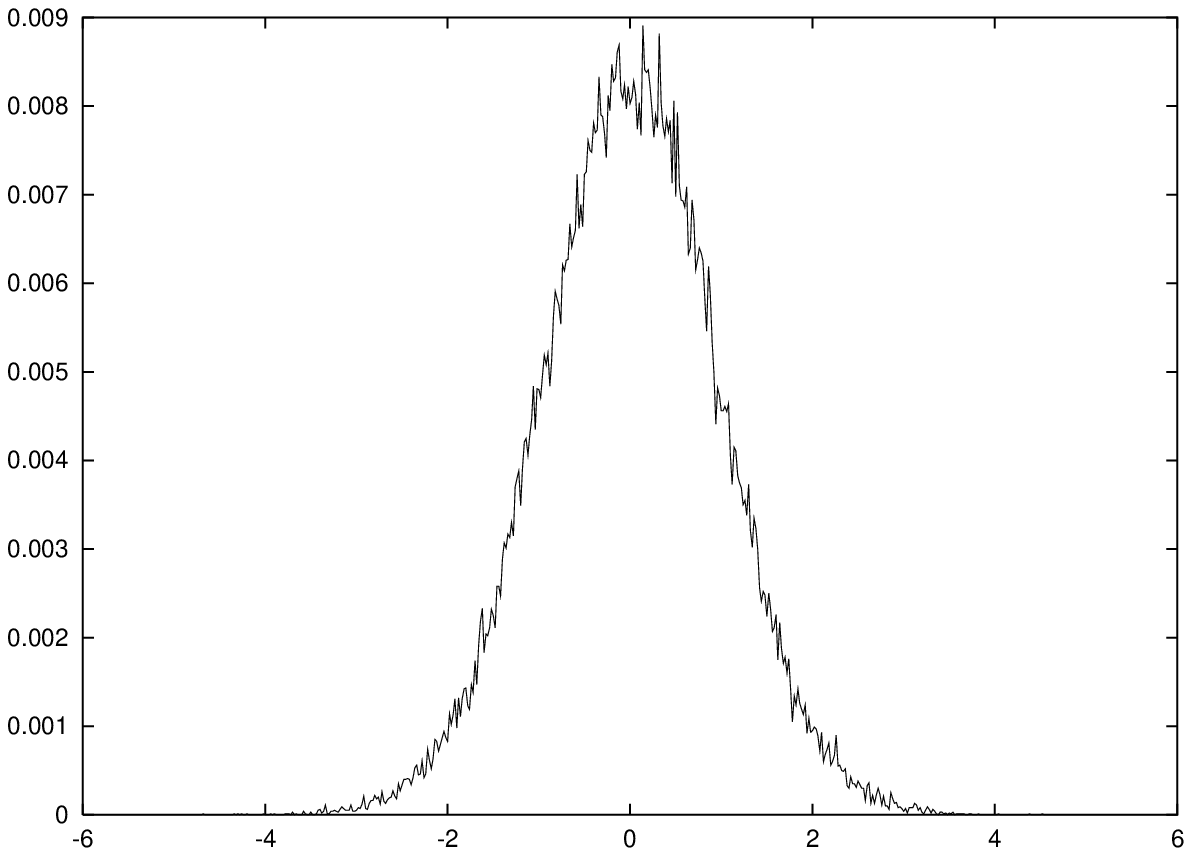}}

\put(42,-8){\here{\small $h$}} \put(122,-8){\here{\small
$h$}}\put(202,-8){\here{\small $h$}}

\put(20,65){\here{$\Phi(h)$}} \put(100,65){\here{$\Psi(h)$}}
\put(180,65){\here{$W(h)$}}

\put(20,145){\here{$\Phi(h)$}} \put(100,145){\here{$\Psi(h)$}}
\put(180,145){\here{$W(h)$}}

\end{picture}
\vspace*{8mm}
 \caption{Effective field distributions
 for `small world' spin glasses, i.e. ferromagnetic rings with random sparse long-range
 bonds with strengths distributed according to
 $p(J_{ij})=p\delta[J_{ij}-J]+(1-p)\delta[J_{ij}+J]$.
 Here we show examples for $c=10$, i.e. well above the percolation
 threshold of the Poissonian graph, in the spin-glass phase, with
 $p=5/8$.
 Top row: $J_0/J=1/2$  and $T/J=1/10$ (here the presence of the ring is important).   Bottom row:
 $J_0/J=1/32$  and $T/J=1/8$ (here the physics is dominated by the sparse long-range bonds).} \label{fig:pop_sg1}
\end{figure}

\begin{figure}[t]
\setlength{\unitlength}{0.6mm}\hspace*{12mm}
\begin{picture}(200,160)
\put(0,80){\includegraphics[height=77\unitlength,width=80\unitlength]{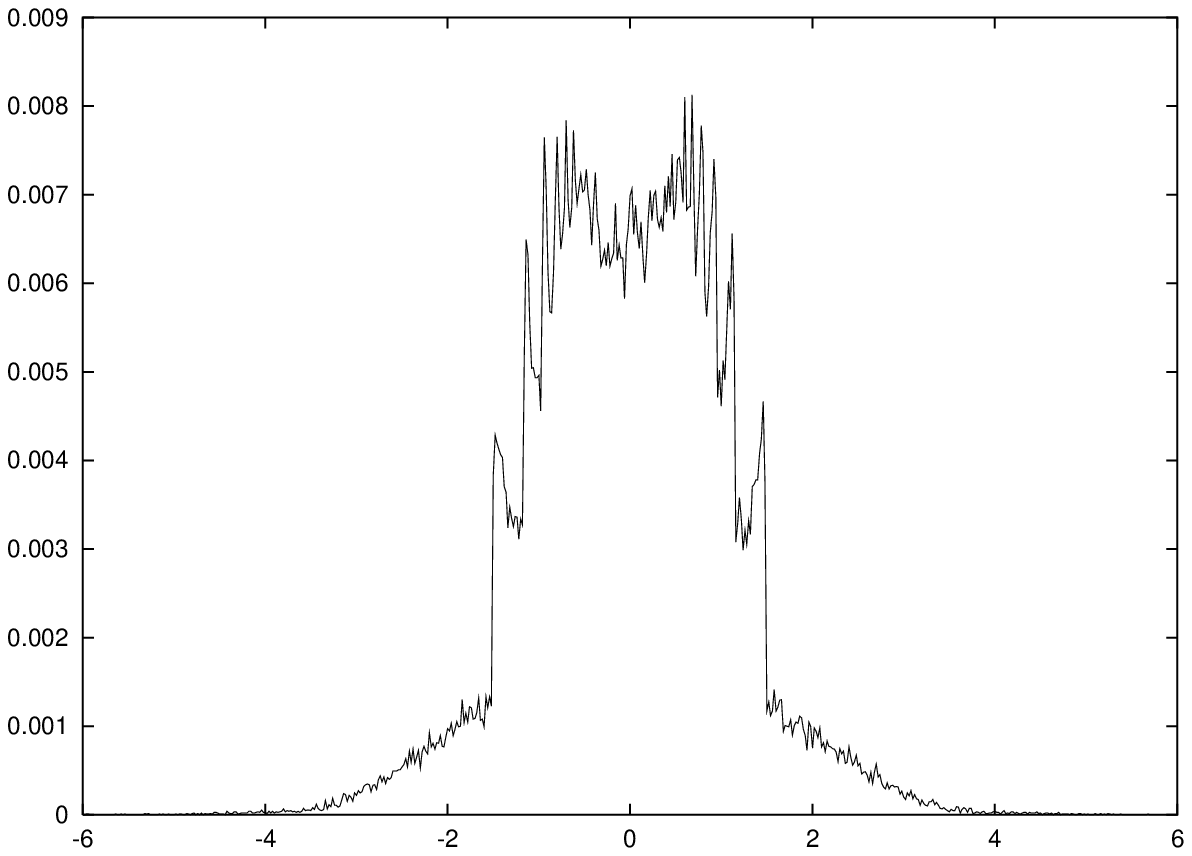}}
\put(80,80){\includegraphics[height=77\unitlength,width=80\unitlength]{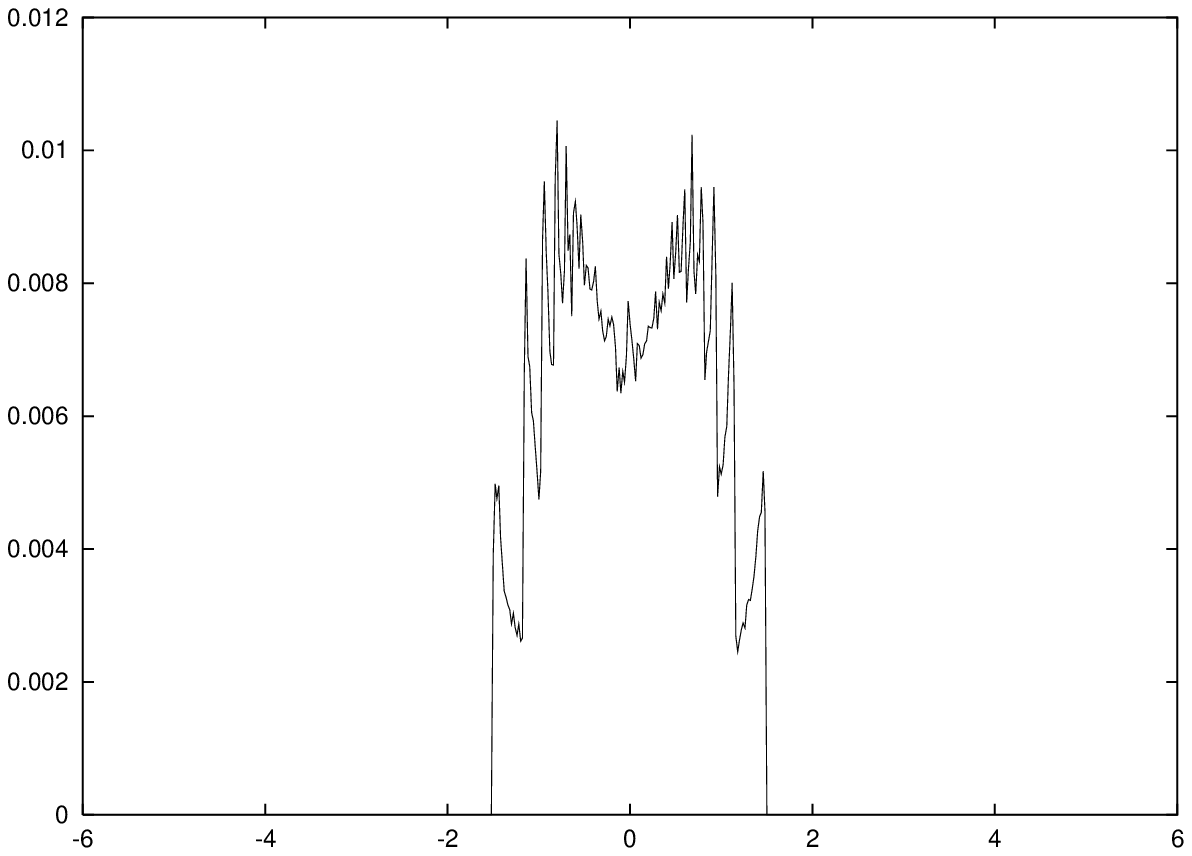}}
\put(160,80){\includegraphics[height=77\unitlength,width=80\unitlength]{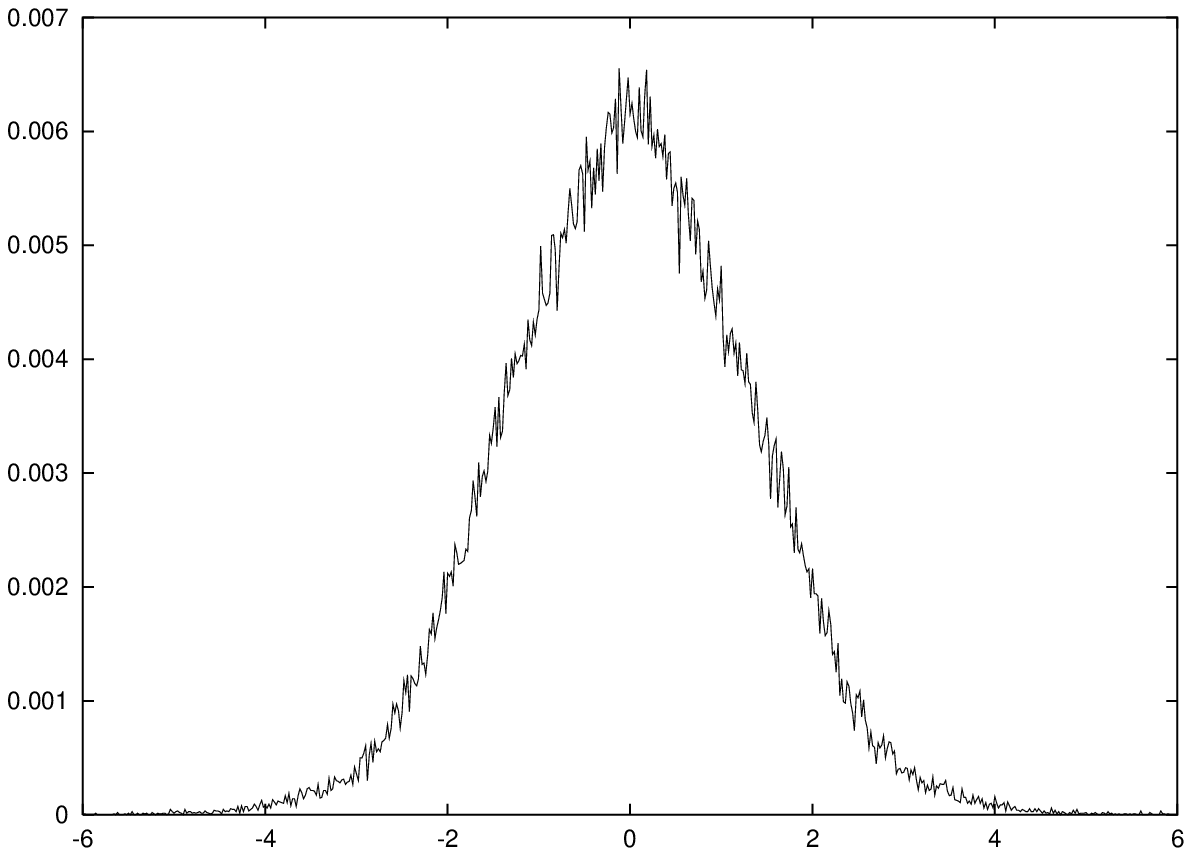}}

\put(0,0){\includegraphics[height=77\unitlength,width=80\unitlength]{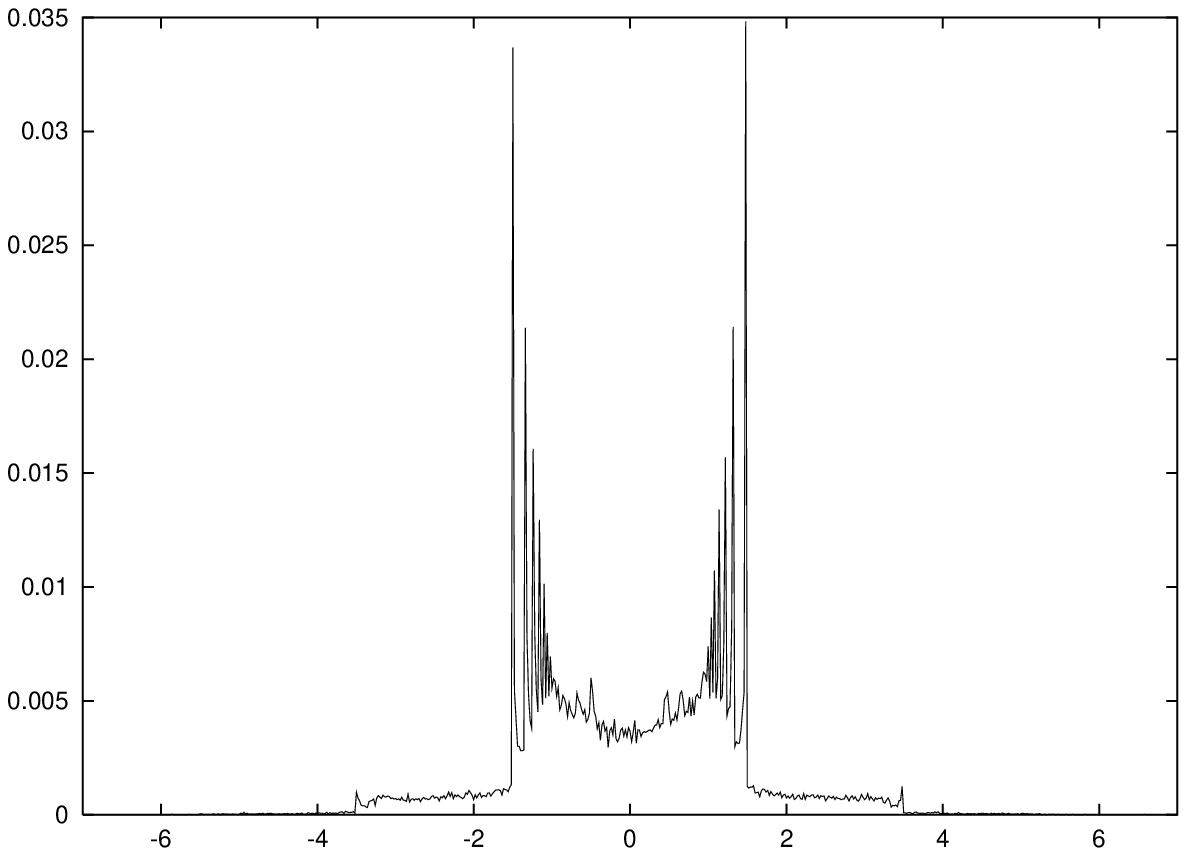}}
\put(80,0){\includegraphics[height=77\unitlength,width=80\unitlength]{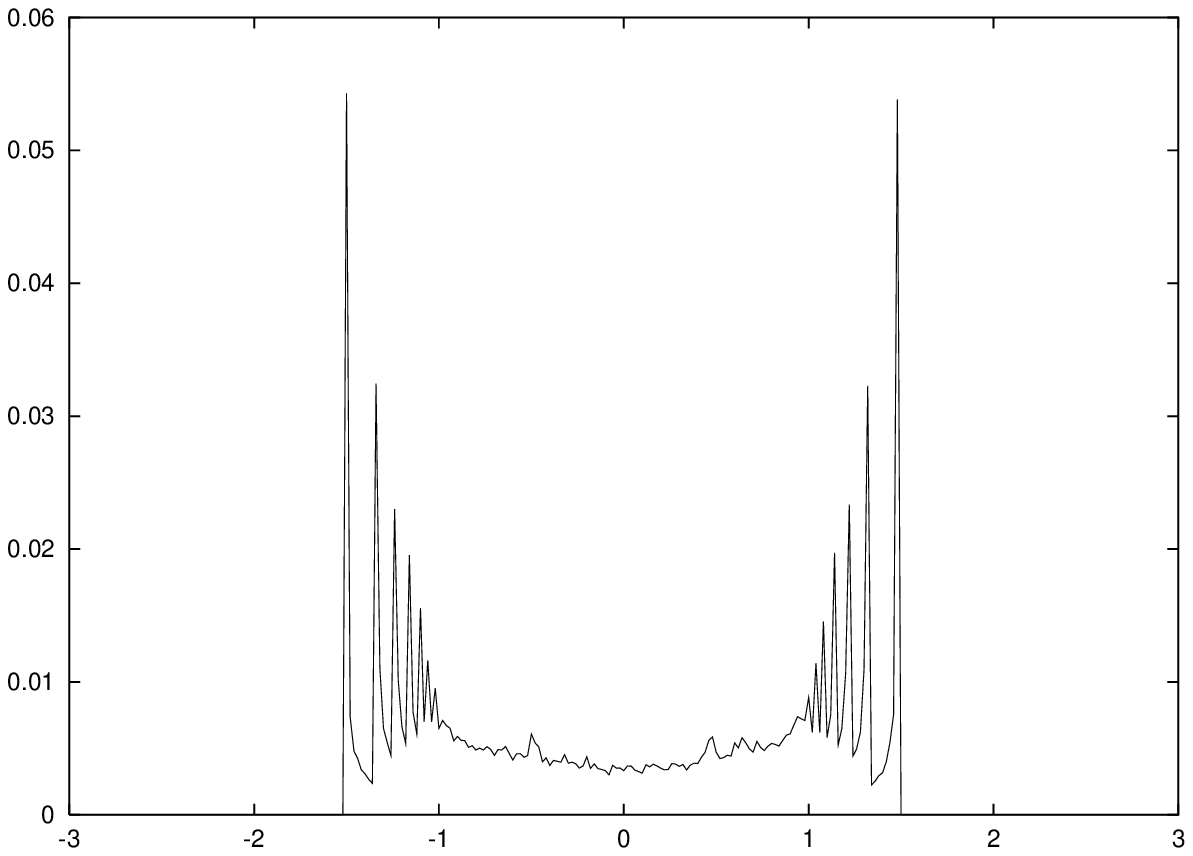}}
\put(160,0){\includegraphics[height=77\unitlength,width=80\unitlength]{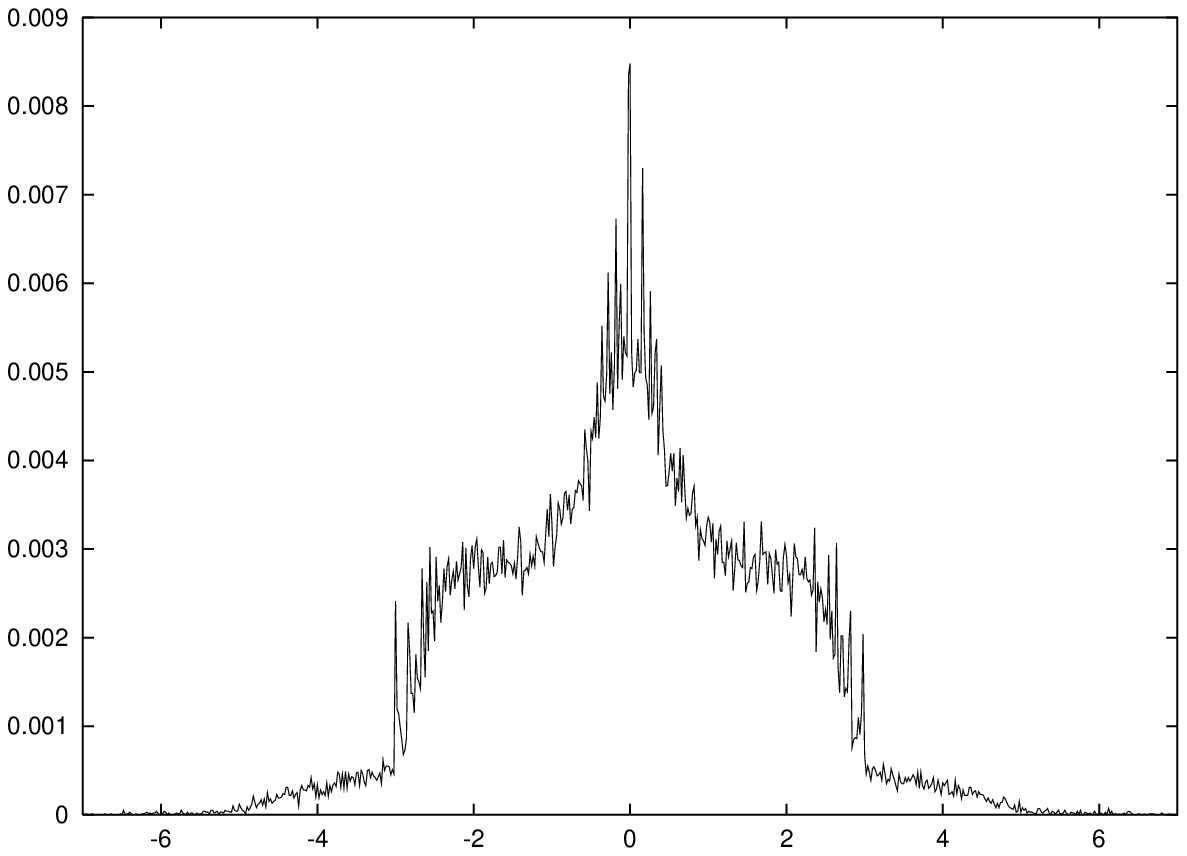}}

\put(42,-8){\here{\small $h$}} \put(122,-8){\here{\small
$h$}}\put(202,-8){\here{\small $h$}}

\put(20,65){\here{$\Phi(h)$}} \put(100,65){\here{$\Psi(h)$}}
\put(180,65){\here{$W(h)$}}

\put(20,145){\here{$\Phi(h)$}} \put(100,145){\here{$\Psi(h)$}}
\put(180,145){\here{$W(h)$}}

\end{picture}
\vspace*{8mm}
 \caption{Effective field distributions
 for `small world' spin glasses, i.e. ferromagnetic rings with random sparse long-range
 bonds with strengths distributed according to
 $p(J_{ij})=p\delta[J_{ij}-J]+(1-p)\delta[J_{ij}+J]$.
 Here we show examples for $c=\frac{1}{2}$, i.e. well below the percolation
 threshold of the Poissonian graph (where the `small world' effect dominates), again in the spin-glass phase, with
 $p=5/8$ and $J_0/J=3/2$.
 Top row: $T/J=1$.   Bottom row: $T/J=1/2$.} \label{fig:pop_sg2}
\end{figure}

\begin{figure}[t]
\setlength{\unitlength}{0.6mm}\hspace*{12mm}
\begin{picture}(200,82)

\put(0,0){\includegraphics[height=77\unitlength,width=80\unitlength]{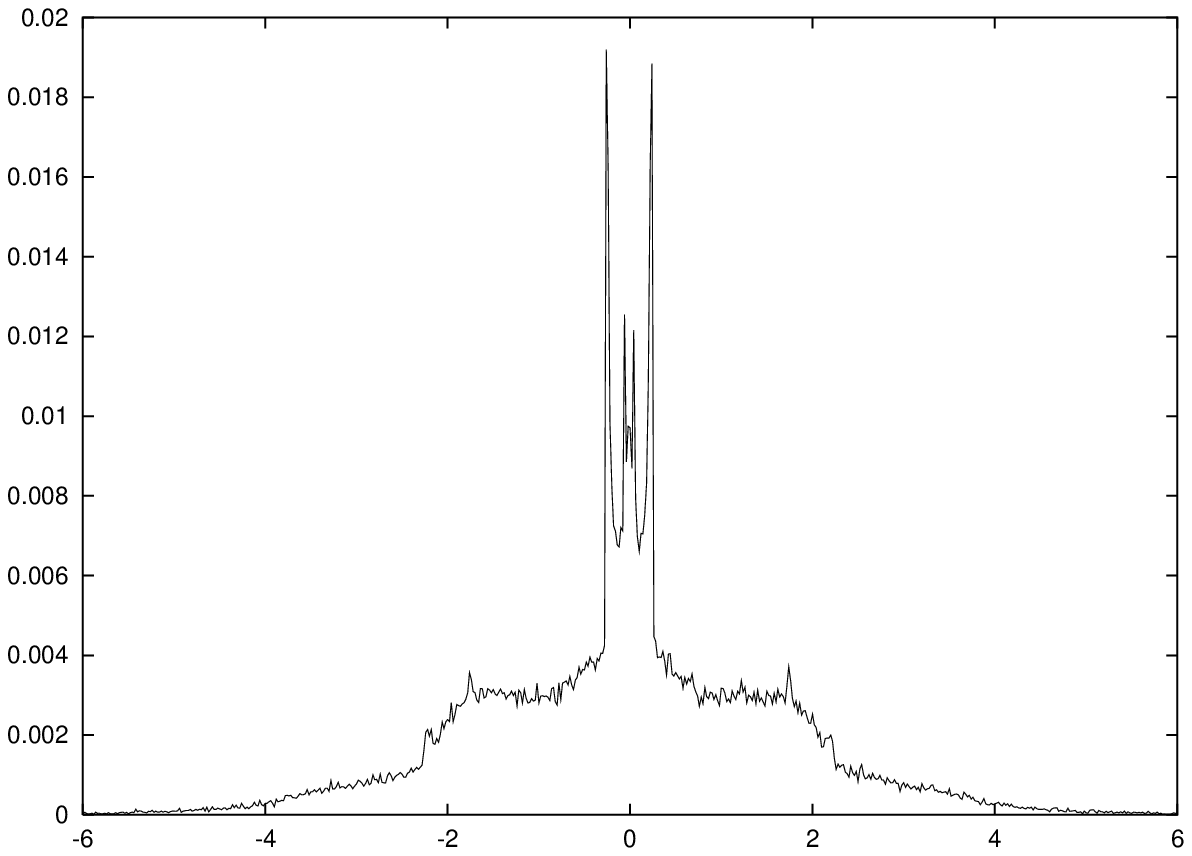}}
\put(80,0){\includegraphics[height=77\unitlength,width=80\unitlength]{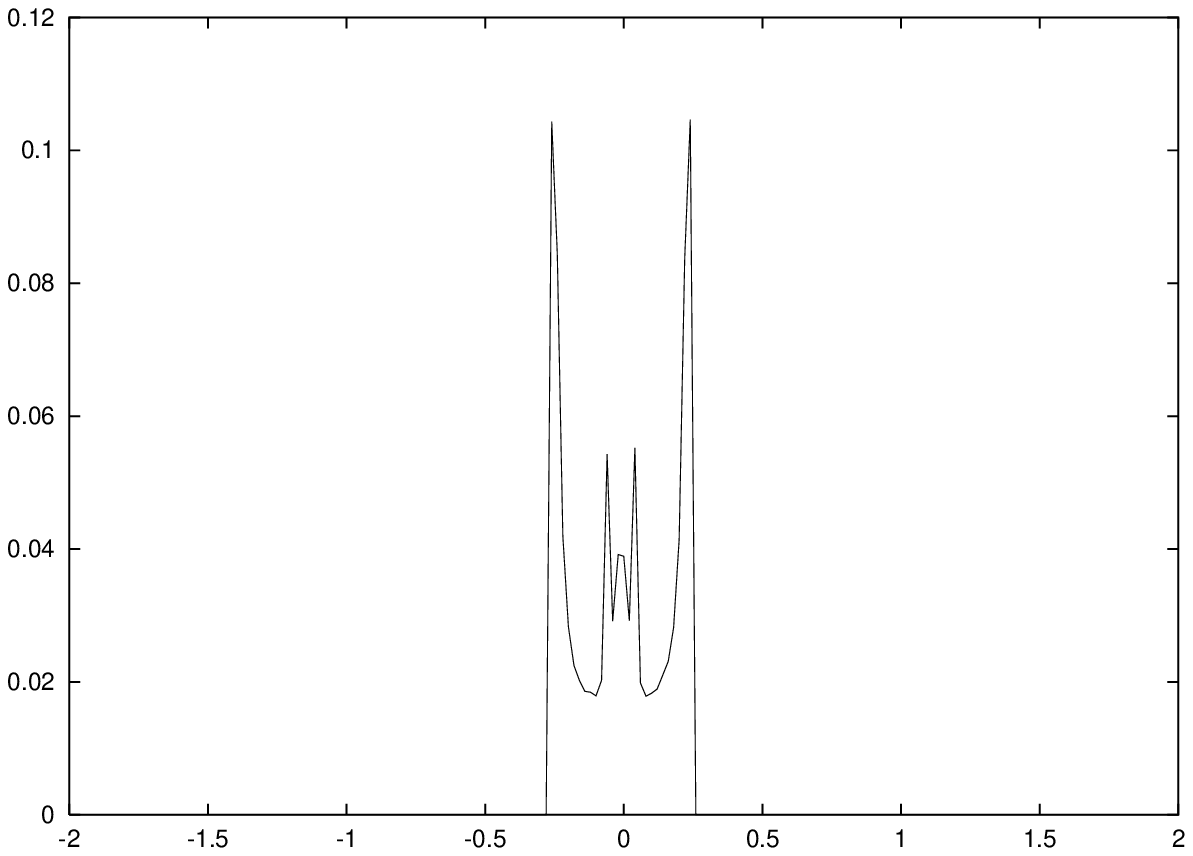}}
\put(160,0){\includegraphics[height=77\unitlength,width=80\unitlength]{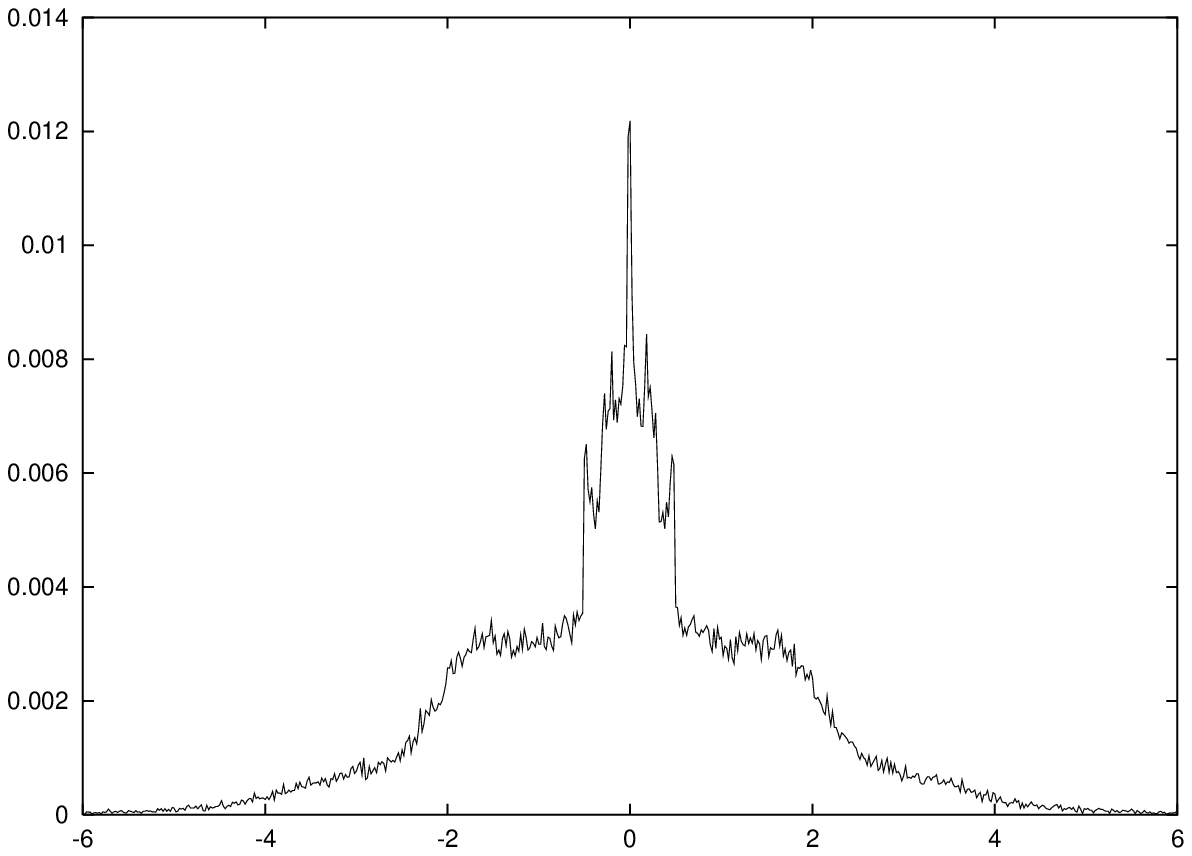}}

\put(42,-8){\here{\small $h$}} \put(122,-8){\here{\small
$h$}}\put(202,-8){\here{\small $h$}}

\put(20,65){\here{$\Phi(h)$}} \put(100,65){\here{$\Psi(h)$}}
\put(180,65){\here{$W(h)$}}

\end{picture}
\vspace*{8mm}
 \caption{Effective field distributions
 for `small world' spin glasses, i.e. ferromagnetic rings with random sparse long-range
 bonds with strengths distributed according to
 $p(J_{ij})=p\delta[J_{ij}-J]+(1-p)\delta[J_{ij}+J]$.
 Here we show examples for $c=2$, i.e. just above the percolation
 threshold of the Poissonian graph, again in the spin-glass phase, with
 $p=3/4$,  $J_0/J=1/16$ and $T/J=1/4$.
} \label{fig:pop_sg3}
\end{figure}

Also for the small world spin-glass  we have solved numerically
the order parameter equations
(\ref{eq:PHI},\ref{eq:PSI},\ref{eq:W}),  resulting in figures
\ref{fig:pop_sg1}, \ref{fig:pop_sg2} and \ref{fig:pop_sg3}. It
will be clear that, especially at low temperatures and around or
below the percolation threshold, the effective field distributions
can acquire highly nontrivial shapes (especially when compared to
the Gaussian effective field distributions which one typically
finds in non-diluted bond-disordered spin models). Furthermore,
increasing the bond strength $J_0$ along the ring has the effect
of changing field distributions from being smooth into more
discretized shapes, similar to what is found in random field and
random bond Ising chains \cite{RF1,RF2,RF3}.

\subsection{Comparison with simulations}

\begin{figure}[t]
\vspace*{5mm}

\setlength{\unitlength}{0.68mm}\hspace*{-10mm}
\begin{picture}(200,80)
\put(109,0){\includegraphics[height=85\unitlength,width=90\unitlength]{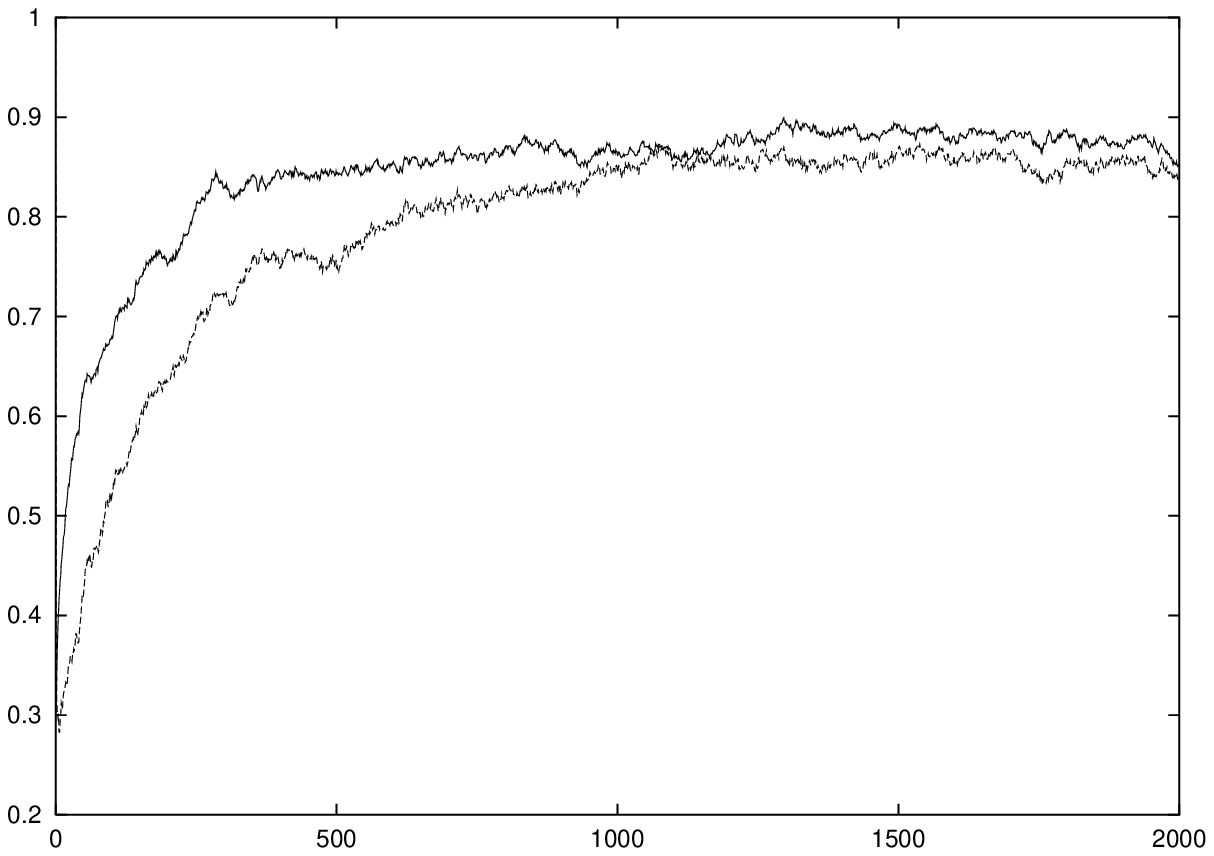}}

\put(157,-8){\here{\small $t$ (iter/spin)}}

\put(95,43){\here{$m,~q$}}
\end{picture}
\vspace*{10mm}
 \caption{Observables $m$ (solid lines) and $q$ (dashed lines)
  as measured during numerical simulations of the `small world'
  spin-glass
 for $N=10,\!000$.
We show equilibration in the ferromagnetic phase, below the
percolation threshold of the Poissonian graph ($p=3/4$, $T=1$,
$c=0.5$, $J_0=2$ and $J=0.5$). Here the theory predicts the
equilibrium values $m\simeq 0.84$ and $q\simeq 0.83$.  Compared to
the small world magnet, equilibration is harder to achieve,
 but the agreement between
theory and simulations is still very good.} \label{fig:sim_sg}
\end{figure}

We have finally  compared the predicted values for $m$ and $q$ of
our solution, as obtained by numerical solution of
(\ref{eq:PHI},\ref{eq:PSI},\ref{eq:W}), followed by evaluation of
(\ref{eq:q_and_m}), with the result of numerical simulation of the
stochastic microscopic dynamics (of the conventional Glauber type)
 with $N=10,\!000$. In contrast to the small world magnet, in the small world
spin-glass equilibration is found to be not only much slower but
also  more subject to finite size fluctuations. However,  the
agreement between theory and experiment is still found to be very
good. Examples are shown in figure \ref{fig:sim_sg}.

\section{Discussion}

\begin{figure}[t]
\setlength{\unitlength}{0.6mm}\hspace*{12mm}
\begin{picture}(200,160)
\put(0,80){\includegraphics[height=77\unitlength,width=80\unitlength]{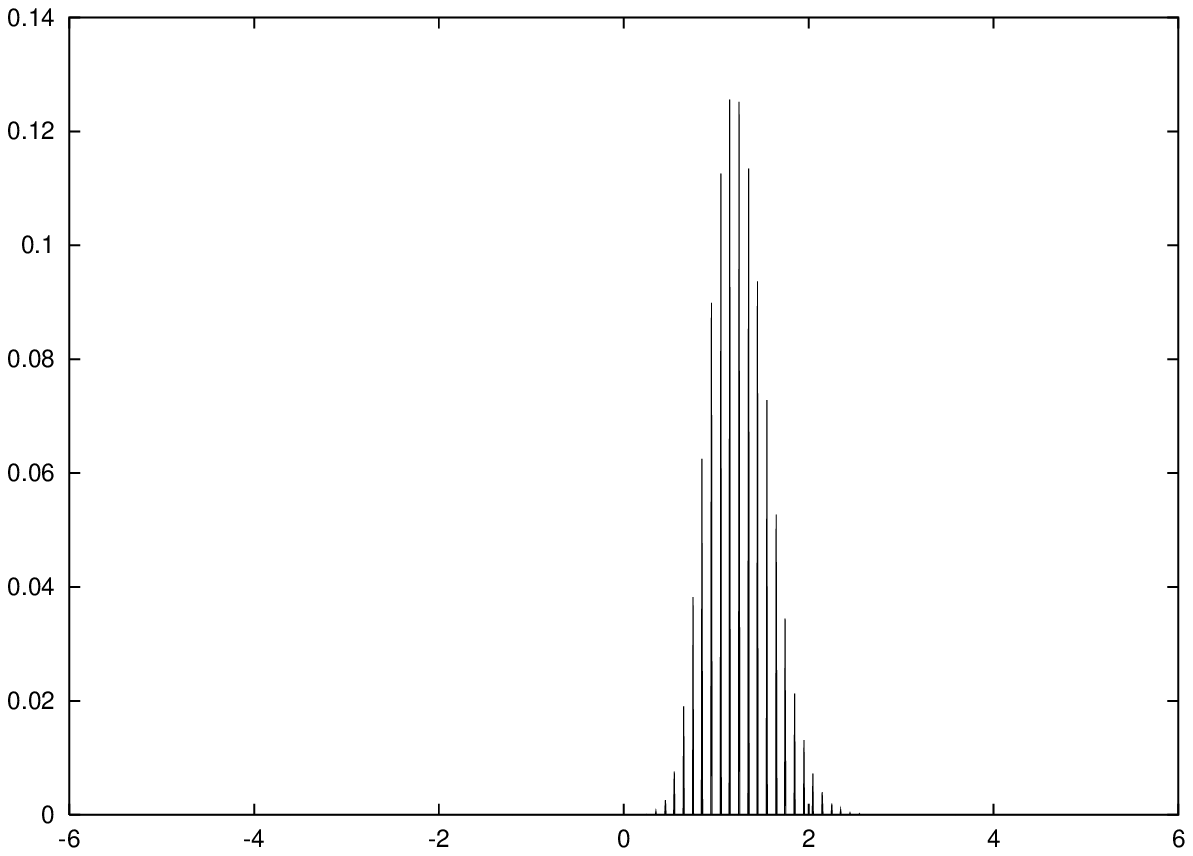}}
\put(80,80){\includegraphics[height=77\unitlength,width=80\unitlength]{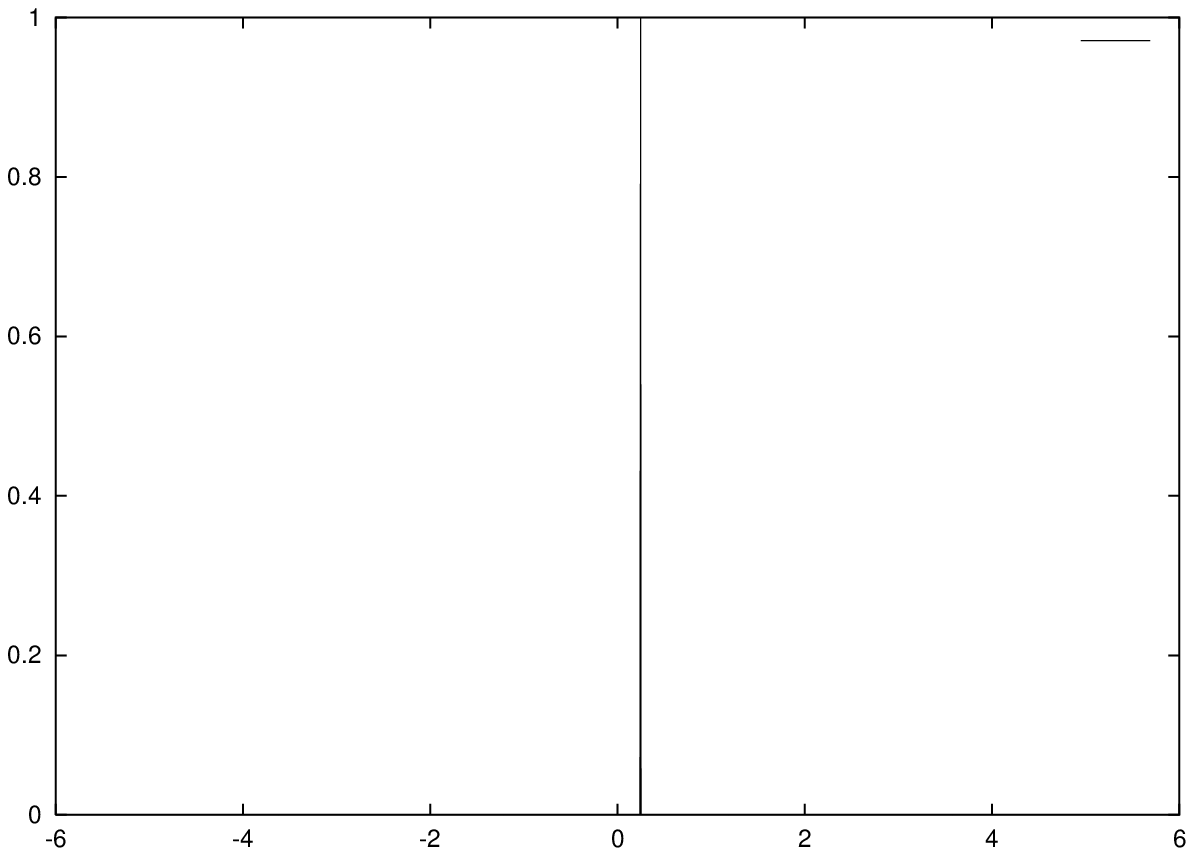}}
\put(160,80){\includegraphics[height=77\unitlength,width=80\unitlength]{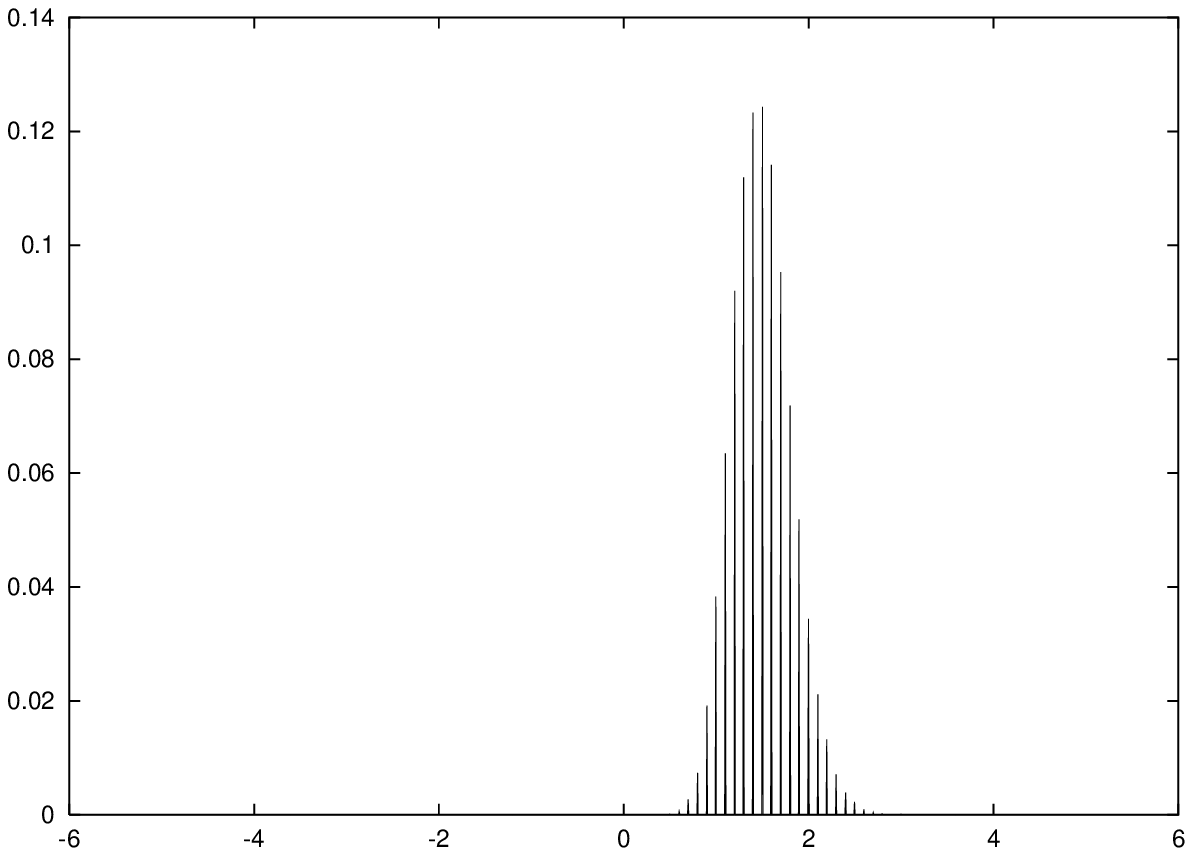}}

\put(0,0){\includegraphics[height=77\unitlength,width=80\unitlength]{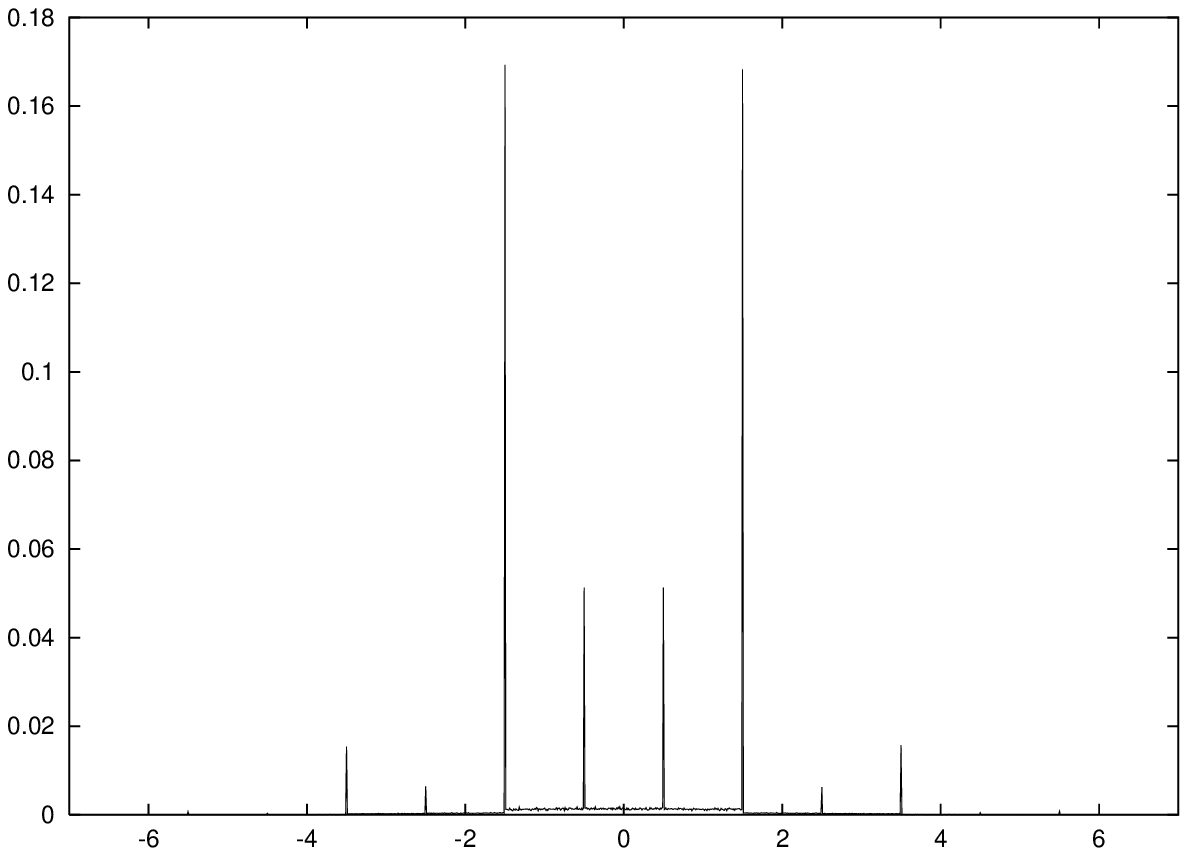}}
\put(80,0){\includegraphics[height=77\unitlength,width=80\unitlength]{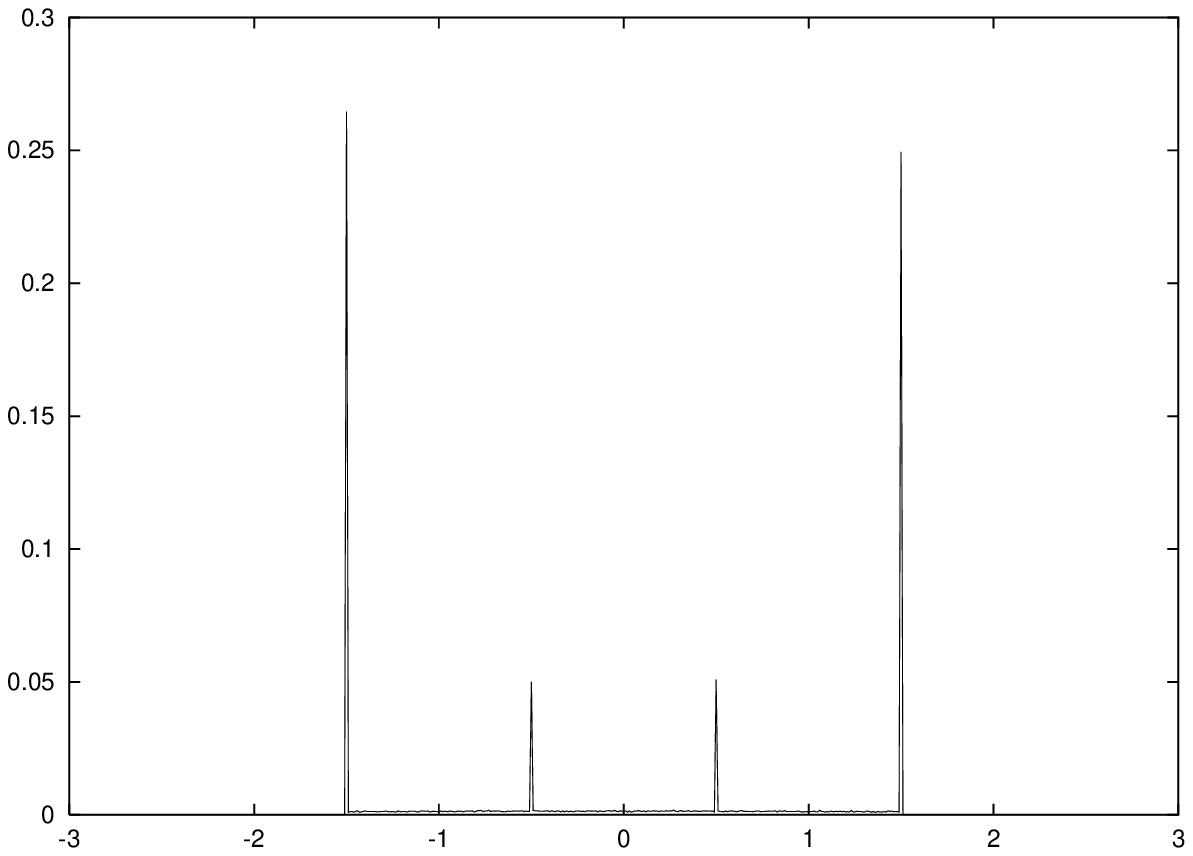}}
\put(160,0){\includegraphics[height=77\unitlength,width=80\unitlength]{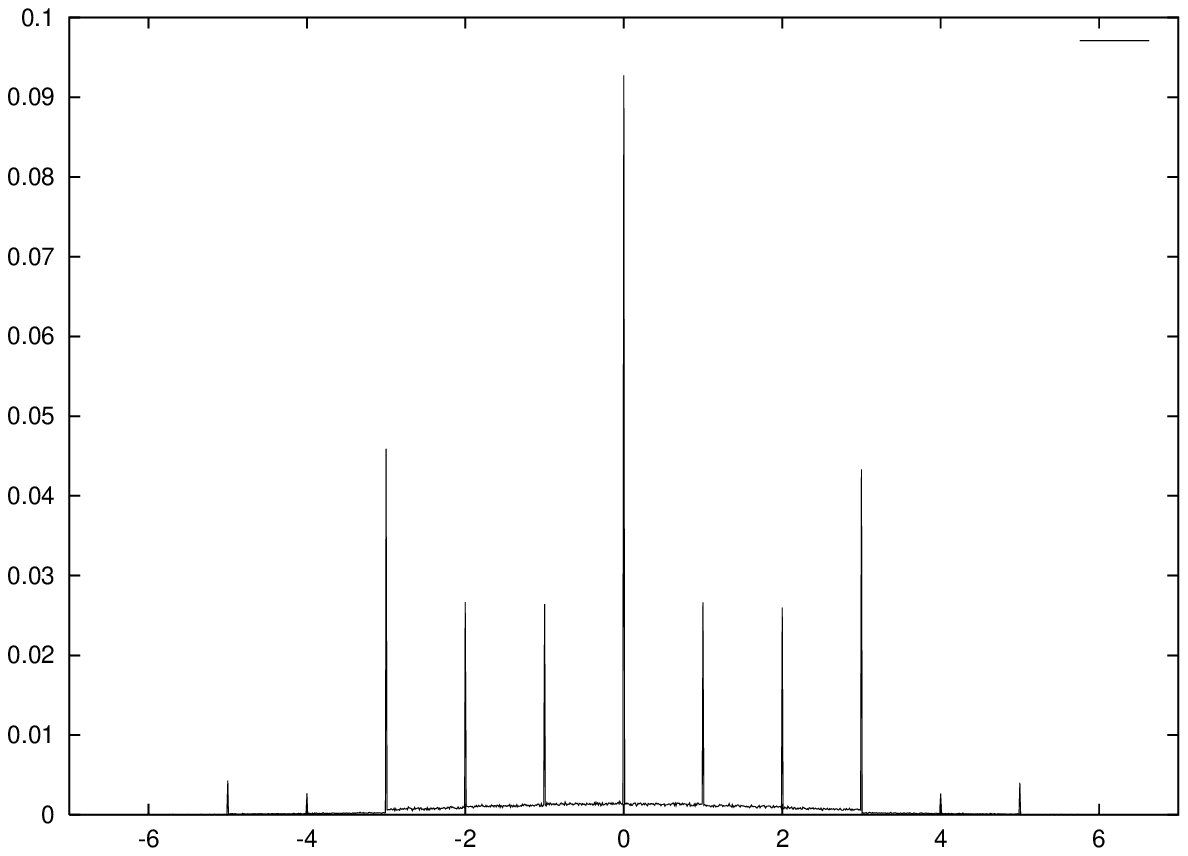}}

\put(42,-8){\here{\small $h$}} \put(122,-8){\here{\small
$h$}}\put(202,-8){\here{\small $h$}}

\put(20,65){\here{$\Phi(h)$}} \put(100,65){\here{$\Psi(h)$}}
\put(180,65){\here{$W(h)$}}

\put(20,145){\here{$\Phi(h)$}} \put(100,145){\here{$\Psi(h)$}}
\put(180,145){\here{$W(h)$}}

\end{picture}
\vspace*{8mm}
 \caption{Effective field distributions at zero temperature,
 for `small world' magnets (i.e. ferromagnetic rings with random sparse
 long-range ferromagnetic
 bonds, top row) and for `small world' spin glasses (i.e. ferromagnetic rings with random sparse long-range
 $\pm J$ random bonds). Parameters for the small world magnet are
 $c=10$, $J_0=\frac{1}{4}$, and $J=1$; for the small
 world spin-glass they are $c=\frac{1}{2}$, $J_0=\frac{3}{2}$,
 $J=1$ and $p=\frac{5}{8}$.
 These data strongly suggest that, whereas the small world magnet
 exhibits zero temperature effective field distributions
 containing only discrete delta-peaks, those found for small world
 spin-glasses have additional continuous pieces. This, in turn,
 might be regarded as evidence for the need to break replica
 symmetry \cite{CavityT0}.
} \label{fig:zeroT}
\end{figure}

In this paper we have analyzed Ising spin models on `small world'
lattices, consisting of a combination of a one-dimensional
ferromagnetic periodic chain and a sparse (finitely connected)
random Poissonian graph with (generally) random bonds, using the
replica method. Upon making a replica symmetric ansatz, we have
been able to diagonalize the relevant replicated transfer
matrices, and thereby obtain explicit  expressions for the
asymptotic free energy per spin and the continuous transitions
away from the paramagnetic phase to either a ferromagnetic or a
spin-glass one. We believe our results to constitute the first
rigorous solution for this type of hybrid random `small world'
Ising spin model (given the RS ansatz), of relevance therefore
both in the context of `small world' models as such, but also in
terms of the methodology used (diagonalization of replicated
transfer matrices, which should have broader applicability). We
have applied our theoretical results to two specific cases: the
`small world' ferromagnet (sparse long-range bonds of uniform
value) and the `small world' spin-glass (sparse random $\pm J$
long-range bonds). For these two specific models we could draw
explicit RS phase diagrams, and also demonstrate analytically the
nontrivial so-called `small world' effect: for any non-zero value
of the bond strength $J_0$ along the chain (however weak), and any
non-zero average Poissonian connectivity $c$ (even $c<1$) there is
always a {\em finite} temperature phase transition to an ordered
state. We have carried out (a limited number of) simulation
experiments, which are found to be in good agreement with our
theory.

As always, many interesting questions remain to be answered and
extensions to be carried out. Firstly, it would be important to
find out whether and how replica symmetry needs to be broken in
the present model. On the one hand, the fact that the SG
transition line in our model is not dependent on $p$ (the
probability for a long-range bond to be $+J$) might be seen as a
reason to doubt the need for RSB, since it suggests that in the
present models frustration does not play a major role (since for
$p=1$ we would have all ferromagnetic long-range bonds, whereas
for $p=-1$ they would be all anti-ferromagnetic). On the other
hand, numerical evaluation of the various effective field
distributions in our model at $T=0$ (see figure \ref{fig:zeroT})
reveals continuous contributions in the small world spin-glass,
which one might take as evidence for RSB. Secondly, one might
wonder about and investigate the extent and role of domain
formation along the spin chain. Obvious generalizations of our
present study would be to other types of long range bond disorder
(e.g. Gaussian), other types of spin variables, non-Poissonian
random graphs (adapting e.g. the methods of
\cite{wong-sherrington88}), to anti-ferromagnetic or disordered
bonds along the one-dimensional chain (where on the basis of
\cite{skantzos-coolen00} one should expect to find first order
transitions), and to include an analysis of correlation and
response functions. In one-dimensional systems correlation
functions are linked to the second-largest eigenvalue of the
transfer matrix. Here we would therefore require further
eigenvalues of the replicated transfer matrix; this calculation
can be carried out using the formalism of \cite{nik-coolen04}, and
will be the subject of a future study.

\section *{Acknowledgment}

This study was initiated during an informal Finite Connectivity
Workshop at King's College London in November 2003. TN, IPC, NS
and BW acknowledge financial support from the State Scholarships
Foundation (Greece), the Fund for Scientific Research (Flanders,
Belgium), the ESF SPHINX programme, and the FOM Foundation
(Fundamenteel Onderzoek der Materie, the Netherlands),
respectively.

\clearpage
\section *{References}

\end{document}